\def\lam{\mbox{$\:\lambda\ $}}

\def\ha{{H$\alpha$}}

\def\kms{\:\rm{\,km\,s^{-1}}}

\def\LUM{\:{\rm ergs\:s^{-1}}}

\def\VEL{\:{\rm km\:s^{-1}}}

\def\etal{{\it et\:al.}}

\def\oiiL{[\ion{O}{2}] $\lambda 3727$}

\def\SiiL{[\ion{S}{2}] $\lambda\lambda 6717, 6731$}
\def\NiiL{[\ion{N}{2}] $\lambda\lambda 6548, 6583$}

\def\sii{[\ion{S}{2}]}
\def\nii{[\ion{N}{2}]}

\def\oiii{[\ion{O}{3}]}

\def\hii{\ion{H}{2}}
\def\hi{\ion{H}{1}}
\def\perpix{\rm pix^{-1}}

\documentclass[12pt,preprint]{aastex}
\begin{document}

% Additional private definitions that appear to work only inside document

\newcommand{\MSOL}{\mbox{$\:M_{\sun}$}}  

\newcommand{\EXPN}[2]{\mbox{$#1\times 10^{#2}$}}
\newcommand{\EXPU}[3]{\mbox{\rm $#1 \times 10^{#2} \rm\:#3$}}  % exponent with units
\newcommand{\POW}[2]{\mbox{$\rm10^{#1}\rm\:#2$}}
\newcommand{\SING}[2]{#1$\thinspace \lambda $#2}
\newcommand{\MULT}[2]{#1$\thinspace \lambda \lambda $#2}
\newcommand{\CHINU}{\mbox{$\chi_{\nu}^2$}}
\newcommand{\vsini}{\mbox{$v\:\sin{(i)}$}}
\newcommand{\LSOL}{\mbox{$\:L_{\sun}$}}

\newcommand{\fuse}{{\it FUSE}}
\newcommand{\hst}{{\it HST}}
\newcommand{\iue}{{\it IUE}}
\newcommand{\euve}{{\it EUVE}}
\newcommand{\einstein}{{\it Einstein}}
\newcommand{\rosat}{{\it ROSAT}}
\newcommand{\chandra}{{\it Chandra}}
\newcommand{\xmm}{{\it XMM-Newton}}
\newcommand{\swift}{{\it Swift}}
\newcommand{\asca}{{\it ASCA}}
\newcommand{\galex}{{\it GALEX}}
\newcommand{\cxo}{CXO}

% End of defining things

\shorttitle{SNRs in M83}
\shortauthors{Blair \etal}

%\slugcomment{Submitted version, 9 July 2012}
\title{
% \today\\
{The Magellan/IMACS Catalog of Optical Supernova Remnant Candidates in M83\footnote{
Based on observations made with  
the 6.5 meter Magellan Telescopes located at Las Campanas Observatory, and NASA's 
Chandra X-ray Observatory.
The ground-based observations were obtained through NOAO
which is operated by Association of Universities for Research in Astronomy, Inc. 
for the National Science Foundation.
NASA's Chandra Observatory is operated by Smithsonian Astrophysical Observatory 
under contract \# NAS83060 and the data were obtained through program GO1-12115.}}}

\author{
William P. Blair\altaffilmark{1,4},
P. Frank Winkler\altaffilmark{2,4}
 and
Knox S. Long\altaffilmark{3,4}
}
\altaffiltext{1}{The Henry A. Rowland Department of Physics and Astronomy, 
Johns Hopkins University, 3400 N. Charles Street, Baltimore, MD, 21218; 
wpb@pha.jhu.edu}
\altaffiltext{2}{Department of Physics, Middlebury College, Middlebury, VT, 05753; 
winkler@middlebury.edu}
\altaffiltext{3}{Space Telescope Science Institute, 3700 San Martin Drive, 
Baltimore, MD, 21218;  long@stsci.edu}
\altaffiltext{4}{Visiting Astronomer, Las Campanas Observatory, La Serena, Chile}
%\altaffiltext{5}{Visiting Astronomer, Gemini South Observatory, La Serena, Chile}

\begin{abstract}
We present a new optical imaging survey of supernova remnants in M83, using data obtained with the Magellan I 6.5m telescope and IMACS instrument under conditions of excellent seeing.  Using the criterion of strong \sii\  emission relative to H$\alpha$, we confirm all but three of the 71 SNR candidates listed in our previous survey, and expand the SNR candidate list to 225 objects, more than tripling the earlier sample.  Comparing the optical survey with a new deep X-ray survey of M83 with \chandra, we find 61 of these SNR candidates to have X-ray counterparts. We also identify an additional list of 46 \oiii -selected nebulae for follow-up as potential ejecta-dominated remnants, seven of which have associated X-ray emission that makes them  strong candidates.  Some of the other \oiii-bright objects could also be normal ISM-dominated supernova remnants with shocks fast enough to doubly ionize oxygen, but with \ha\ and \sii\ emission faint enough to have been missed.  A few of these objects may also be \hii\ regions with abnormally high \oiii\ emission compared with the majority of M83 \hii\ regions,  compact nebulae excited by young Wolf-Rayet stars, or even background AGN.  The  supernova remnant \ha\ luminosity function in M83 is shifted a factor of $\sim 4.5\times$ higher than for M33 supernova remnants, indicative of a higher mean ISM density in M83.  We describe the search technique used to identify the  supernova remnant candidates and provide basic information and finder charts for the objects.  

Subject Headings: galaxies: individual (M83) -- galaxies: ISM  -- supernova remnants

\end{abstract}

\section{Introduction \label{sec_intro}}

M83 (NGC\,5236) is a classic grand-design SAB(s)c spiral galaxy with a starburst nucleus, active star formation along the arms, and prominent dust lanes \citep{elmegreen98}.  It has played host to six recorded supernovae (SNe) in the past century, second in number only to NGC\,6946 with nine. In this paper we adopt a distance to M83 of 4.61 Mpc \citep{saha06}, and so 1\arcsec\,=\,22 pc.  With its proximity and nearly face-on orientation,  M83 affords the most detailed view of any galaxy where such active star formation and destruction are taking place.  The integrated effects of this active star formation process are manifest through the generally high metallicity and the chemical abundance gradients measured by spectroscopy of \hii\ regions across the $\sim$10\arcmin\ diameter bright optical disk  \citep{{bres02}, {pily06}, {pily10}}.  A fainter and much more extended disk is seen in \hi\  and in GALEX ultraviolet imaging data \citep{{hucht81}, {thilker05}, {bigiel10}}.

% The distance modulus is 28.32.

Of M83's historical SNe, the three with spectroscopically determined types are either Ib or II, both of which result from core-collapse of massive stars \citep{barbon99}.  
Simple extrapolation from the recent past thus leads us to expect that there must have been dozens of core-collapse SNe in M83 within the past millennium, and many more older supernova remnants (SNRs) as well, since expectations are that SNRs remain visible for tens of thousands of years, depending on local conditions in the interstellar medium (ISM) around each object.

In the first systematic attempt to identify SNRs in M83, \citet[][henceforth BL04]{blair04} found 71 SNR candidates  based on the ratio of \SiiL\ lines to \ha\ in CCD imagery.  This criterion has long proven to be a versatile technique for identifying evolved ISM-dominated SNRs, where the SN blast wave propagating through the surrounding ISM produces strong \sii\ and other low-excitation lines in the cooling and recombination zone behind the shock.  Typically, SNRs display \sii:\ha\ ratios $\gtrsim 0.4$, while photoionized nebulae have \sii:\ha\ $\lesssim 0.2$\, providing good separation of these different classes of objects.  In some galaxies, this gap in the ratio is blurred, causing  potential confusion in the application of this criterion, but BL04 found the \sii:\ha\ ratio to work well for M83.   BL04 spectroscopically confirmed 23 (out of 25 observed) of their ISM-dominated SNR candidates, providing at least partial confirmation and vetting of the candidate list from their imaging survey.

BL04 also carried out a separate search for \oiii-bright nebulae in order to search for ejecta-dominated SNRs, similar to Cas A \citep{kirshner77, fesen01} or G292+1.8 \citep{ghav12} in our Galaxy or 1E 0102-7219 in the Small Magellanic Cloud \citep{dopita84, blair00}.  Such a search is realistic in M83 because of the high metallicity; \hii\ regions in M83 are so metal-rich that they can readily cool themselves, reducing their effective ionization levels.  Hence, the vast majority of M83 \hii\ regions have low \oiii:\ha\ ratios.  Very early-type Wolf-Rayet (W-R) stars may also have sufficent ionizing potential to overcome the high abundances and produce extended \oiii\ emission \citep{naze03}.  Hence, extended \oiii-emitting nebulae in M83 are expected to be either normal ISM-dominated SNRs with high enough shock velocities to excite \oiii, potential ejecta-dominated SNRs, or possible nebulae excited by early-type W-R stars.  Young, ejecta-dominated SNRs could be too small for us to resolve (e.g., at the distance of M83, Cas A would have a diameter of only 0.15\arcsec).  Such objects might be confused with planetary nebulae (PNe), which typically have strong \oiii\ emission; however, all the known, ejecta-dominated SNRs have moderately strong X-ray emission, while the X-ray flux from even the brightest  known PNe  would fall far below the detection threshold at the distance of M83.
The most extreme objects found by BL04  had \oiii\lam 5007:\ha\ ratios of only 0.8-0.9, far lower than extreme ejecta-dominated remnants like Cas A, G292.0+1.8, or 1E 0102-7219.  The lack of success in finding such objects was attributed to the mediocre seeing (typcially 1.3\arcsec, or $\sim$30 pc) in their data, which could have smeared out small angular size \oiii-dominated nebulae. 

\citet[henceforth D10]{dopita10} recently reported the results from an imaging study, carried out  with the new Wide Field Camera 3 (WFC3) on the {\it Hubble} Space Telescope, of a single $162\arcsec \times 162\arcsec$ field in M83 that includes the complex nuclear region and part of one spiral arm.  They identified 60 SNR candidates that are relatively bright in both \sii\ and \oiiL\ relative to \ha, only 12 of which had been identified previously by BL04.  In addition, D10 have identified six (slightly) extended nebulae with \oiii\ emission that they suggested may be young, ejecta-dominated SNRs, one compact \oiii\ nebula with a corresponding X-ray source that is almost certainly an ejecta-dominated SNR, and the likely counterpart to SN1968L, which occurred deep within the complex starburst nuclear region.

We have carried out a new survey of the entire bright optical disk of M83 using the 6.5m Magellan-I telescope and the IMACS instrument in imaging mode. We used narrow-band imaging to study the population of SNRs and other nebulae, and broad-band imaging to investigate the stellar populations from which these arise.  Here we  report the results of our search of these data for ISM-dominated SNRs as well as an initial attempt to identify the expected population of young SNRs arising from core-collapse SNe. The next section describes the data and data processing, while \S3 discusses the identification of candidate SNRs and their properties.   In \S4 we provide a brief discussion and comparison with previous results, and a summary follows in \S5.

\section{Observations and Data Reduction \label{sec_obs}}

%In this section we provide the observational details for imaging and spectroscopy of M83 and the SNR candidates. 

%\subsection{Magellan/IMACS Imaging}

We observed M83 from the 6.5m Magellan-I (Baade) telescope at the Las Campanas Observatory (LCO), using the IMACS instrument in its  f/4.3 direct-imaging mode, on the nights of 2009 April 26 and 27 (UT), using time that was assigned through the NOAO time allocation process.   The IMACS camera has a $4\times 2$ mosaic of $2048\times 4096$ chips and covers a 14\arcmin\ square field, encompassing all of the bright disk of M83, at a scale of $0.11\arcsec\,\perpix$.  We carried out imaging in \oiii, \ha, and \sii\ emission lines, plus green and red continuum bands for subtracting stars to better reveal the nebular emission. 
The \ha\ and continuum filters are all standard IMACS ones, but for \oiii\ and \sii, for which there are no appropriate filters at LCO, we were able to borrow filters from CTIO, where they are normally used with the Mosaic camera on the 4m Blanco telescope.  We obtained multiple dithered images of M83 in each filter, at positions separated by $\sim 18\arcsec$ in both N-S and E-W directions in order to fill in the chip gaps and to reduce pixel-to-pixel variations in the final mosaicked images.  Seeing throughout the run was superb, generally 0.4\arcsec\,-\,0.5\arcsec.  The characteristics of the filters and observations are summarized in Table \ref{im_log}.   We note that the \ha\ filter also likely passes some portion of  \NiiL, as described below.   While \nii\ contributes a variable component to the flux through this filter, for simplicity we shall refer to emission measured through it simply as ``\ha."  Also, when we refer to \oiii\ below, it just means the stronger $\lambda$5007 line of the doublet.

The standard data reduction included line-by-line overscan correction and trimming, bias subtraction and flat fielding using dome flats, and was carried out in IRAF\footnote{IRAF is distributed by the National Optical Astronomy Observatory, which is operated by the Association of Universities for Research in Astronomy, Inc., under cooperative agreement with the National Science Foundation.} using the {\tt mscred} package.   Placing the data on a standard astrometric system was complicated by the fact that there are few well measured stars over the bright inner region of M83.  We used stars from the UCAC2 and USNO-B1.0 catalogs \citep{zacharias04, monet03}, selecting ones with small positional uncertainties and checking visually to eliminate a few background galaxies included in the catalogs.  This typically gave 50 - 100 stars on each of the eight CCDs.  Even so, there were not enough stars (especially near the center of M83) to reliably determine the distortion corrections at all the rotation angles used, so we used a tangent-plane projection and fit the positions separately for each chip using simple bilinear fits.   This produced excellent results, with typical RMS errors of $< 80$ mas, comparable to the uncertainties in the catalogs themselves.   We then used the tasks {\tt mscimage} and  {\tt mscstack} to reproject and combine all the images onto a standard system.   Typical FWHM for stars in the combined images is $\sim 0.5\arcsec$, barely larger than on the individual frames.  

We matched the point spread functions of the combined continuum images to the emission-line ones (green to \oiii, and red to \ha\ and \sii), scaled, and subtracted to remove most of the stellar continuum as well as individual bright stars.   This simple procedure is remarkably effective at revealing faint nebulosity.  For convenience we also used the IRAF task {\tt imsurfit} to fit  a planar surface to the outer regions of the combined images and subtracted the fit surface to set the residual background sky level to near zero.  In Fig.~\ref{fig_overview}, we show a single full-frame image of the reduced \ha\ data (before continuum subtraction) with a reference grid that will be used later in this paper.  Looking ahead, this figure also uses colored symbols to show the overall spatial distribution of various categories of objects identified below, as described in the figure caption.

To achieve absolute flux calibration, we observed seven spectrophotometric standards from the list of \citet{hamuy92} over a range of airmass and processed these identically to our M83 images.  Photometry of these stars gave a conversion factor between count rate and flux in each filter, with an rms dispersion of $<2\%$ for each of the emission lines.   We then applied the mean conversion factors to the continuum-subtracted images, allowing us to measure integrated fluxes for each object or region of interest.  However, in comparing our results to BL04 and to other available calibrated data sets for M83, it became clear that our \ha\ fluxes were being underestimated relative to the other lines.  Upon further research, it appears that the  effective bandpass of the \ha\ filter in the converging f/4.3 beam of the Magellan telescope was shifted from its nominal value to the blue more than expected, as indicated in Table \ref{im_log}.  This moved the \ha\ line onto the edge of the filter bandpass and thus reduced the measured flux levels.  

To quantify this effect and correct our \ha\ fluxes, we obtained the calibrated images of M83 reported by \citet{herrmann08} and measured 15 objects in common between the two data sets.   We also extracted \ha\ fluxes for the same objects from the SINGS survey data on M83 \citep{meurer06}.  We found excellent agreement between the two comparison data sets, but as expected our derived fluxes were systematically low.  Because the scatter in these comparisons was small, any potential impact of the systematic radial velocity gradient across the M83 disk from southwest to northeast \citep[$\sim 200 \kms$, as measured by][in H I and CO]{crost02}, was  deemed small enough to be neglected.  Comparing to these data sets, we derived a correction factor of 1.45 for the Magellan \ha\ fluxes, which has been applied to the values as listed in Tables \ref{table_s2} and \ref{table_o3}.  
%Because the widths of the other filters are somewhat broader than the \ha\ filter, those images are unaffected by this problem and no correction is needed.

To complicate matters further, as mentioned previously the \ha\ filter passed some emission from \NiiL.  This causes two problems.  First, the strength of the \nii\ lines relative to \ha\ approximately follows the relative abundances, and so in spiral galaxies with abundance gradients, one expects the  \nii:\ha\ ratios to vary with galactocentric distance; if \nii\ contaminates the \ha\ image, this will produce somewhat  smaller image-derived \sii:\ha\ ratios nearer the center where the \nii\ lines are stronger.   Second and  perhaps more importantly,  the \nii\ lines are also much stronger in SNRs than in \hii\ regions (see for example BL04), which also tends to reduce the observed image-derived \sii:\ha\ ratio.  In our case,  the blueward shift of the ``\ha" filter bandpass mentioned above shifts the stronger \nii\ line at 6583 \AA\  almost entirely out of the filter bandpass, thus reducing this contamination, but   the weaker \nii\ line at 6548 \AA\ lies fully within the filter and will cause a modest and variable contamination of the derived \ha\ fluxes.    However, our results will show that this contamination effect is not severe enough to significantly affect our ability to distinguish SNRs  from \hii\ regions.

% \citet{crost02} indeed show that in CO emission, their is a roughly 180 $\VEL$  gradient across the face of M83,  from $\sim$600 $\VEL$ in the NE to $\sim$420 $\VEL$ in the SW (corresponding to a shift of $\pm$2 \AA.      

As a final step and to aid in the SNR searches described below, we used the aligned, reduced, and continuum-subtracted images to produce \sii:\ha\ and \oiii:\ha\ ratio images.  We first set a floor for each image at a level that cuts out sky noise. We then adjusted this floor level to an intermediate gray display level of 0.4 for each of the ratios.  Objects with lower values for the ratio will show as black in the image display, and ones with a higher ratio will show as white.  Displaying these ratio images alongside the actual data guides the eye to regions of interest that can then be inspected in more detail.  The one caveat to this technique is that the faintest emission-line objects can be impacted adversely if their flux levels are close to the selected threshold.  However, our search will necessarily be incomplete for the faintest sources anyway. 

We have used the flux-calibrated images to extract flux information in \ha, \sii, and \oiii, and then derive image-based ratios in \sii:\ha\ and \oiii:\ha\ (with the caveat caused by the \nii\ contamination). We accomplished this by using the SAOimage ds9  display program  \citep{joye03} to create  tightly defined circular regions around each object of interest, the diameters of which are recorded in column 5 of Tables \ref{table_s2} and \ref{table_o3}.  
 Even though backgrounds had been subtracted from the images, we defined background regions near each object to allow local or overlying diffuse emission to be properly removed. Background regions were typically larger than the object regions to average out the noise and provide improved background subtraction for each object.  
 %After extracting the background-corrected total counts from each object in each data set, the flux conversion factors from the standard star measurements were used to convert to total fluxes in $\FLUX$.  
The derived background-subtracted fluxes and ratios are listed in Tables \ref{table_s2} and \ref{table_o3}. 

\section{Identifying SNRs in M83 \label{sec_find}}

In the following sub-sections we discuss the criteria applied for finding SNR candidates in M83 and then discuss the extraction of flux information for the candidates.

\subsection{ISM-dominated SNRs\label{ism_snrs}}

The now-classic application of the \sii:\ha\ criterion in finding SNRs is that objects with \sii:\ha\ $\ge$ 0.4 are considered shock heated (and thus SNR candidates), based on expectations from radiative shock models \citep{hartigan87, allen08}.  However, it is often the case that the real power of this diagnostic arises from the fact that most photoionized regions have ratios well below 0.4, and so there is a clear gap separating \hii\ regions from SNRs.  This is not always the case, especially as one pushes to lower surface brightnesses, and so in galaxies such as NGC~7793 and NGC~300 (Sculptor group spirals; see \citet{blair97}), the gap in ratio space  is populated by objects, causing significant confusion  for objects near the dividing line.  Happily, in M83 the gap seems to be quite well-defined and there are relatively few objects with ratios near the dividing line.  This is important since the contamination of our \ha\ image by strong and spatially variable \nii\ emission, as described above, could in principle have confused the situation much more than it apparently does.

We performed a new search for ISM-dominated SNRs using the following methodology.  Continuum-subtracted emission-line images and selected continuum band images were loaded into separate frames in the SAOimage ds9 image display, along with the \sii:\ha\ ratio image described above.  Displaying all of the frames simultaneously as a tiled grid, we then used the zoom and align functions in ds9 to systematically inspect and compare the appearances of each region of the M83 field.  We displayed identifiers showing the BL04 candidates, which could then be independently vetted in our new data as we searched for new SNR candidates.  (All but a few objects were---see below.)  We have also obtained a new listing of X-ray sources in M83 as part of a deep (730 ks) \chandra\ survey in progress by ourselves and others (Long et al. 2012b, in preparation).  A region file containing this source list was also displayed so that any X-ray detections of the objects could be noted.  As an example of the search process, Fig.\  \ref{fig_example} shows a $\sim$50\arcsec\ region northwest of the nucleus, enlarged sufficiently to show detail.  For display purposes, we show a three-color version of the subtracted emission line images in the left panel, and the aligned \sii:\ha\ ratio image in the middle panel.  The right panel shows the green continuum image so that the stellar component in the region can be judged separately.  Details are given in the figure caption.

\begin{figure}
\plotone{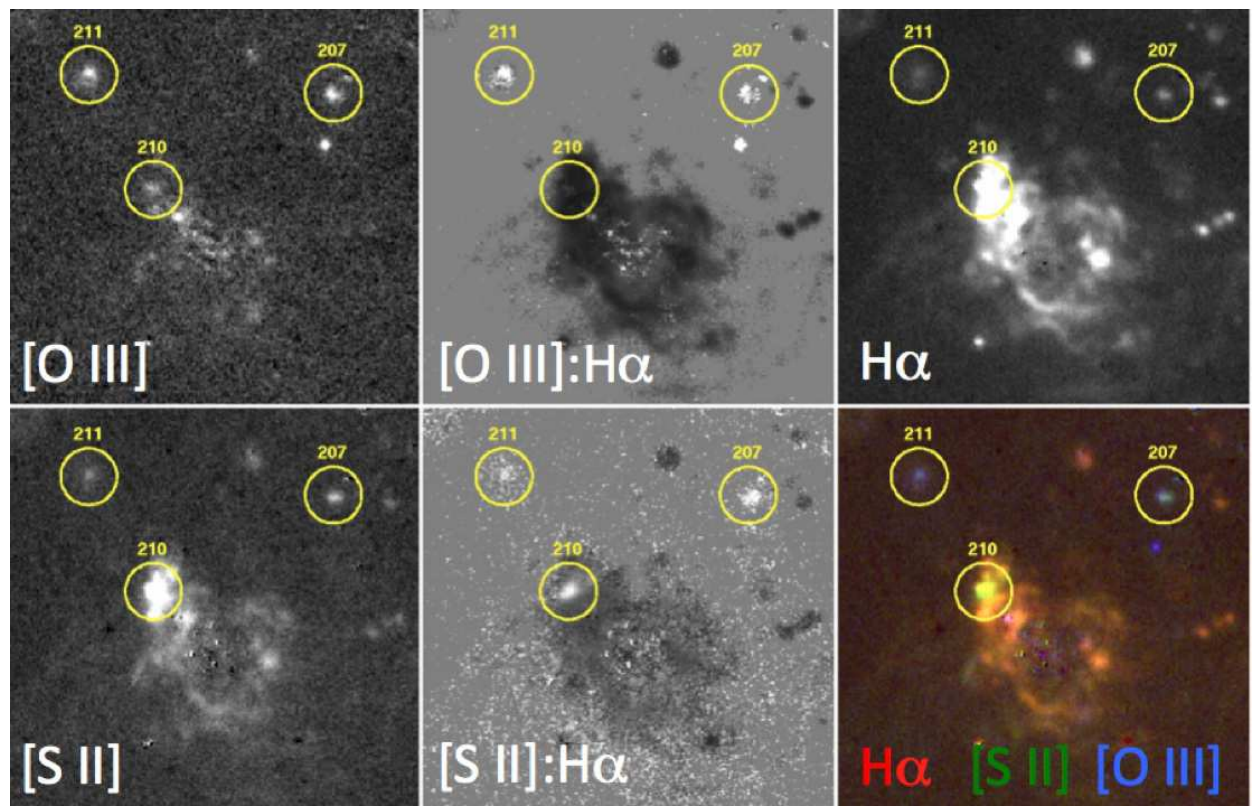}
\caption{This six-panel figure shows another example of the emission-line diagnostics for finding normal ISM-dominated SNR candidates.   A 27\arcsec\ square region is shown with a giant \hii\ region and three normal SNR candidates.  The top row shows the subtracted \oiii, the \oiii:\ha\ ratio image and the subtracted \ha, while the bottom three panels continue with the \sii, the \sii:\ha\ ratio image, and a three-color representation of the emission line data, as indicated on the panels.  The upper two green circles indicate objects 207 and 211, which were identified because of high \sii:\ha\ ratio but which also have significant \oiii\ emission, giving them a bluish color in the lower right panel.  The third green circle indicates object 210, which was identified by the \sii:\ha\ ratio despite being buried in a region of bright \hii\ emission.  This object may have faint \oiii\ emission, but not enough to make it register as interesting in the \oiii:\ha\ panel.  Some faint stellar subtraction residuals are visible in the center of the \hii\ region complex. The bright dot just below object 207 in the [O~III] panels is a likely PN. \label{fig_example2}}
\end{figure}

In Fig.\  \ref{fig_example2}, we show a smaller region including a bright \hii\ region to demonstrate how the \sii:\ha\ technique can  work even in confused regions if the source is sufficiently bright.  The \sii:\ha\ ratio panel clearly shows a region of high ratio buried within a bright clump of emission in the shell of the \hii\ region, indicated by the lower green circle.  Toward the top of the figure,  two other green circles indicate additional SNR candidates identified because of their high \sii:\ha\ ratio, but these two also show moderately strong \oiii\ emission as well, thus modifying their appearance in the color display.  As described in the caption, a PN is also present in this Figure for comparison.

As can be judged from these figures, the ratio image was of particular importance for drawing the eye to regions of potential interest, but it was not applied blindly.  Stellar residuals in the images could cause false high ratio regions that could be readily diagnosed by looking at the individual images.  Regions of very low surface brightness could cause false positives (or false negatives) as the errors in the pixel-by-pixel ratio technique become larger.  Each candidate was carefully inspected in the individual images, and the faintest objects had to be judged to have a degree of morphological integrity to be considered a viable SNR candidate.  While any such search involves the application of a certain amount of judgment, especially for the faintest objects, every attempt was made to be as systematic and consistent as possible.  Even so, the search will still be incomplete at the faintest levels or for faint objects in regions of highest contamination by bright and complex \hii\ emission.  Each new candidate was marked with a ds9 region indicator for later tracking purposes and comparison with other data sets.

Because of the exceptional seeing and the corresponding data quality, we have been able to greatly expand the SNR candidate list in M83.  We have identified 157 new ISM-dominated SNR candidates using the \sii:\ha\ technique, and have confirmed that all but three of the previous 71 BL04 candidates satisfy our identification criteria.  As a result, we have increased the total number of ISM-dominated SNRs and SNR candidates to 225.  We find that 61 of the SNR candidates correspond with \chandra\ X-ray sources in the deep survey data of Long et al. (2012b). Also, of the 40 non-nuclear SNR candidates identified by D10 in one \hst/WFC3 field, we have independently identified 25.  Cross references to previously-identified SNR candidates are indicated in the last column of Table \ref{table_s2}.

There are at least two aspects to the success we have had here in identifying SNR candidates---both attributable in large part to the excellent seeing.  One is that we have been able to go deeper in exposure, and thus see fainter objects.  Comparing the fluxes for the faintest objects reliably detected in BL04 and our Table  \ref{table_s2}, the current survey goes a factor between 4 and 5 deeper.  Even more important, however, is our success in identifying candidates in relatively confused regions on the outskirts of \hii\ complexes and star forming regions (see Fig.\  \ref{fig_example}) that were badly confused in earlier data.  Indeed, there are numerous examples in the list of SNR candidates clustered around the outside edges of giant \hii\ complexes.  Many of the BL04 candidates are extended sources, and some show resolved morphological structure at the resolution of these data, with shells or arcs visible where only a diffuse patch of emission had been seen previously.  

Three of the BL04 candidates, objects 26, 27, and 65 in the BL04 catalog, do not appear to have high enough \sii:\ha\ ratios to be cataloged as SNR candidates in these new data; all are relatively faint and of low surface brightness and may be impacted by the limitations of the ratio image, as discussed earlier.   In the absence of optical spectra for these objects in BL04, we have removed these three objects from the current list.  One BL04 object, \#30 in their list, is a compact, high surface brightness object with strong \oiii\ emission but weak \ha\ and \sii.  This object was removed from the ISM-dominated SNR list, but was moved to the \oiii-selected object list, which we discuss in the following sub-section.  This object is a strong candidate to be an ejecta-dominated SNR.

We list  the 225 ISM-dominated SNR candidates (BL04 plus newly selected objects), ordered by R.A., in Table \ref{table_s2}.  A new running identification number is provided in the first column, and cross references to previous names or identifications are provided when applicable. We also have calculated and tabulate the galactocentric distances for the objects, based on a central coordinate of RA(J2000) = 13:37:00.95, Dec(J2000) = $-$29:51:55.50, from the NASA Extragalactic Database, an inclination of 24 degrees, and  a major axis position angle of 45 degrees \citep{talbot79}.  The D(ext) column in the table shows the diameter of the circular region extracted for the flux measurements discussed below. This value can be used as a surrogate for the object sizes, although it will necessarily be an overestimate, especially for the smallest objects, since it was sized to include all of the flux from each object.  We expect detailed morphological and accurate size information for these objects will be forthcoming from an upcoming Cy19 \hst/WFC3 observing program.

\subsection{[O~III]-Selected Objects\label{o3_snrs}}

The reasoning and strategy for an [O~III]-based search was outlined in \S1.  Because the mean ionization level of photoionized (\hii\ region) gas in M83 is generally low, the \oiii:\ha\ ratio becomes a useful diagnostic for SNRs. Generally speaking, the observed \oiii:\ha\ ratios in M83 \hii\ regions are $\sim$0.1 -- 0.2 over most of the bright disk where the mean abundance levels are super-solar.  Hence, small angular size nebulae (up to $\sim$5\arcsec\ or about 110 pc) that emit significant \oiii\ emission are of immediate interest for one of two reasons.   First, they may be normal ISM-dominated SNRs with shock velocities in excess of $\sim$100$\VEL$, which will have \oiii\ emission comparable to \ha, thus providing some additional diagnostic power for these objects (e.g. that their shock velocities are above this threshold).  The second and more interesting possibility, especially if the objects are of small angular size, is that a high \oiii:\ha\ ratio, even in the absence of strong \sii,  could indicate an ejecta-dominated  young SNR (also called O-rich SNRs) similar to Cas A in our Galaxy.  The situation is more nuanced than for the ISM-dominated remnants, however, because there is no set value of the ratio that provides a physically-determined threshold; we adopt a pragmatic threshold of \oiii:\ha\ $\ge$0.4 (more than twice the typical \hii\ region value) to identify objects of possible interest.  For spatially unresolved nebulae, however, there is an unavoidable confusion with PNe, which are also strong \oiii-emitters and of which hundreds are known in M83 \citep{herrmann08}. A caveat is that other sources of hard photoionization,  such as X-ray binaries \citep{pakull86} and early-type W-R stars, are hot enough to potentially excite \oiii\ emission in compact but slightly extended nebulae, especially in lower abundance situations such as the Magellanic Clouds \citep{naze03}  In principle, background QSO/AGN could also be present in projection. This caveat will be discussed further below.  

Thus, to compile a list of \oiii-selected objects of interest above and beyond the ISM-dominated SNR sample, we apply the following search criteria: (1) Spatially unresolved \oiii\ sources with X-ray counterparts and elevated \oiii:\ha\ ratio, or (2) spatially resolved nebulae with elevated \oiii:\ha\ ratio.  The former are strong candidates to be young SNRs, since any X-ray emission from PNe at the distance of M83 would be undetectable.  For the latter objects, since no PN should be spatially resolved in our data, any extended nebula with enhanced \oiii:\ha\ ratio should indicate either shock heating and/or enhanced O abundances, or one of the hard photoionization options discussed above.  Here again, the presence of soft X-ray emission would almost certainly confirm such objects as SNRs, but since not all SNRs are detected in X-rays at the distance of M83, the absence of X-ray detection is inconclusive.  We note that any young, ejecta-dominated SNRs with X-ray emission below the detection limit of the \chandra\ survey ($\sim5 \times 10^{35} ~  \LUM$) would not be separable from PNe without follow-up spectroscopy, and would be missed in our survey.  However, the known ejecta-dominated SNRs all have X-ray emission well above this threshold.

\begin{figure}
\epsscale{0.8}
\plotone{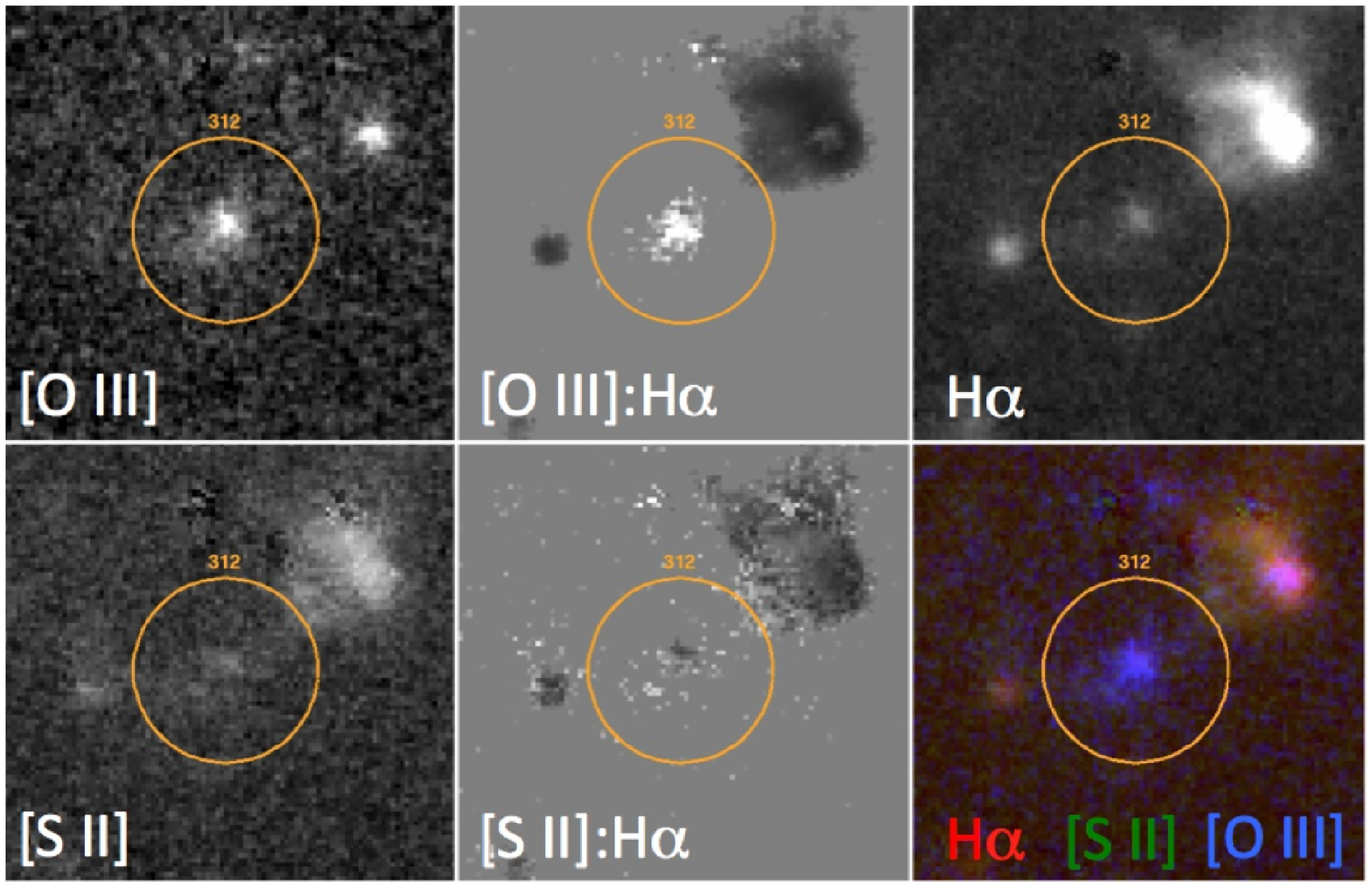}
%\vspace{2mm}
\plotone{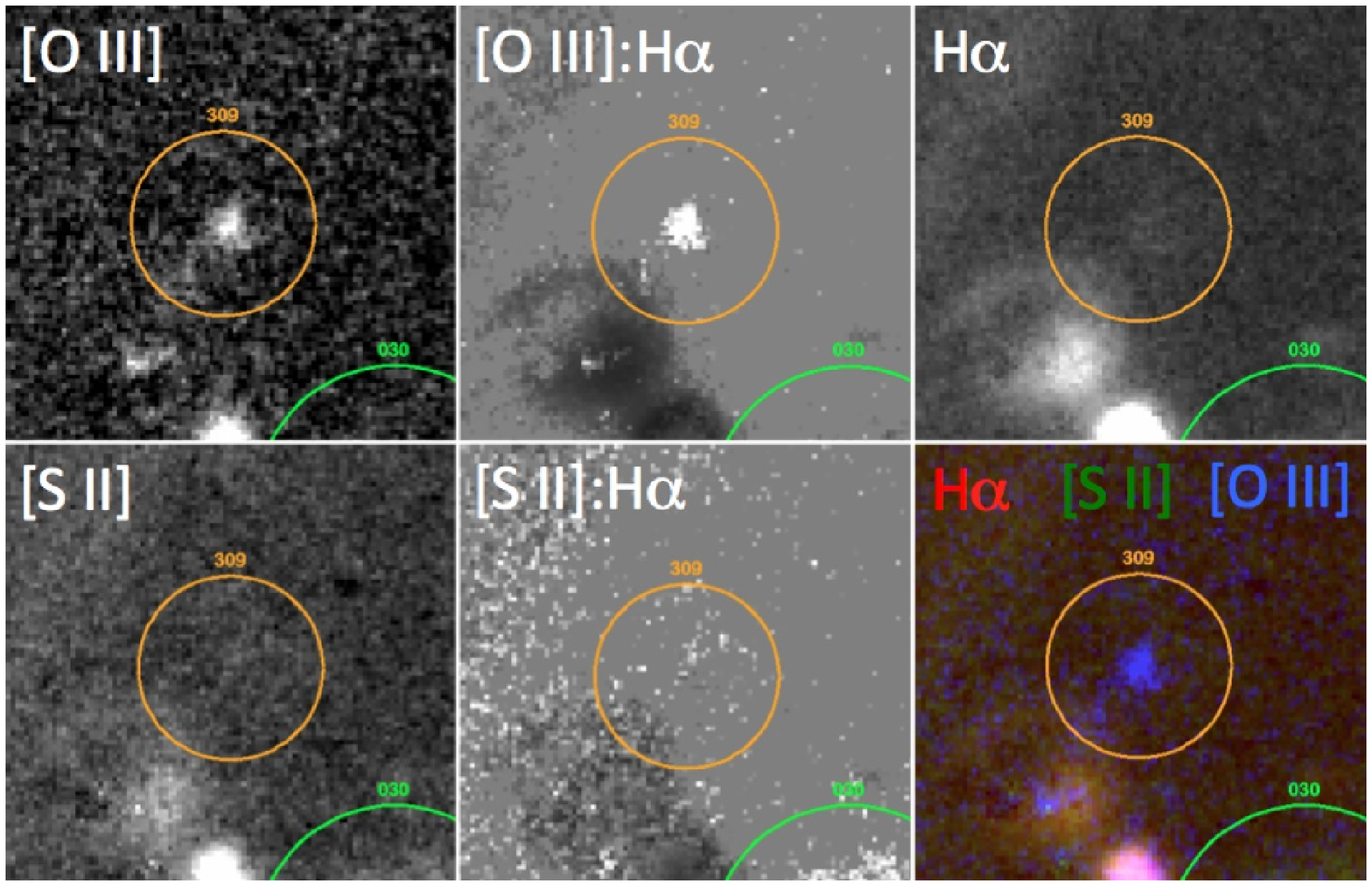}
\caption{Six-panel figures showing two representative examples of \oiii-selected objects. The top example is object 312, and probably represents a normal ISM SNR that was identified because of its relatively strong \oiii.  Faint \ha\ and \sii\ are visible, but were too low surface brightness to register as interesting in the \sii:\ha\ ratio image and so the object was missed.  The bottom example shows object 309, a likely new ejecta-dominated SNR.  This object has \oiii\ emission without detectable \ha\ or \sii\ emission, and has a coincident strong \chandra\ X-ray source.  The region shown for each object is just under 9\arcsec\ square.  \label{fig_example3}}
\end{figure}

As with the ISM SNR search described above, the \oiii:\ha\ ratio image was displayed in conjunction with aligned continuum-subtracted emission-line and continuum comparison images. Region files were displayed that identified the existing ISM-dominated SNR candidates, and also the \chandra\ point source list.  Using the same grid and procedure as described above, the data were once again inspected by eye.  Hundreds of point-like \oiii-emitters with elevated ratios were visible in the data, but we selected only those handful of objects where  an X-ray counterpart was also present.  The vast majority of the objects without X-ray emission are almost certainly PNe.  In the outer parts of the galaxy, we found many of the PNe identified by \cite{herrmann08}, but their survey largely avoided the inner part of the galaxy.  We find many additional uncatalogued PNe in  the inner galaxy as well.  Many extended \oiii-emitters with elevated ratios aligned with ISM-dominated SNR candidates already found above, but a few dozen additional objects of interest were identified.  Two example \oiii-selected objects are shown in Fig. \ref{fig_example3}.

In all, 46 additional objects of interest were identified in this search, as summarized in Table \ref{table_o3}. This Table is ordered on R.A. and a running number starting with ``301" to separate them from the ISM SNRs.   As with the normal SNR candidates, galactocentric distances are also listed for these objects in Table \ref{table_o3}, as described above for the ISM-dominated SNRs.  We list extraction size indicators for each object primarily to show whether the candidate is extended well above the PSF or not.  Again, for most objects, anticipated \hst\ imaging data will provide much better size information, as already shown by D10 for one WFC3 field.

\section{Discussion \label{sec_discussion}}

\begin{figure}
\plotone{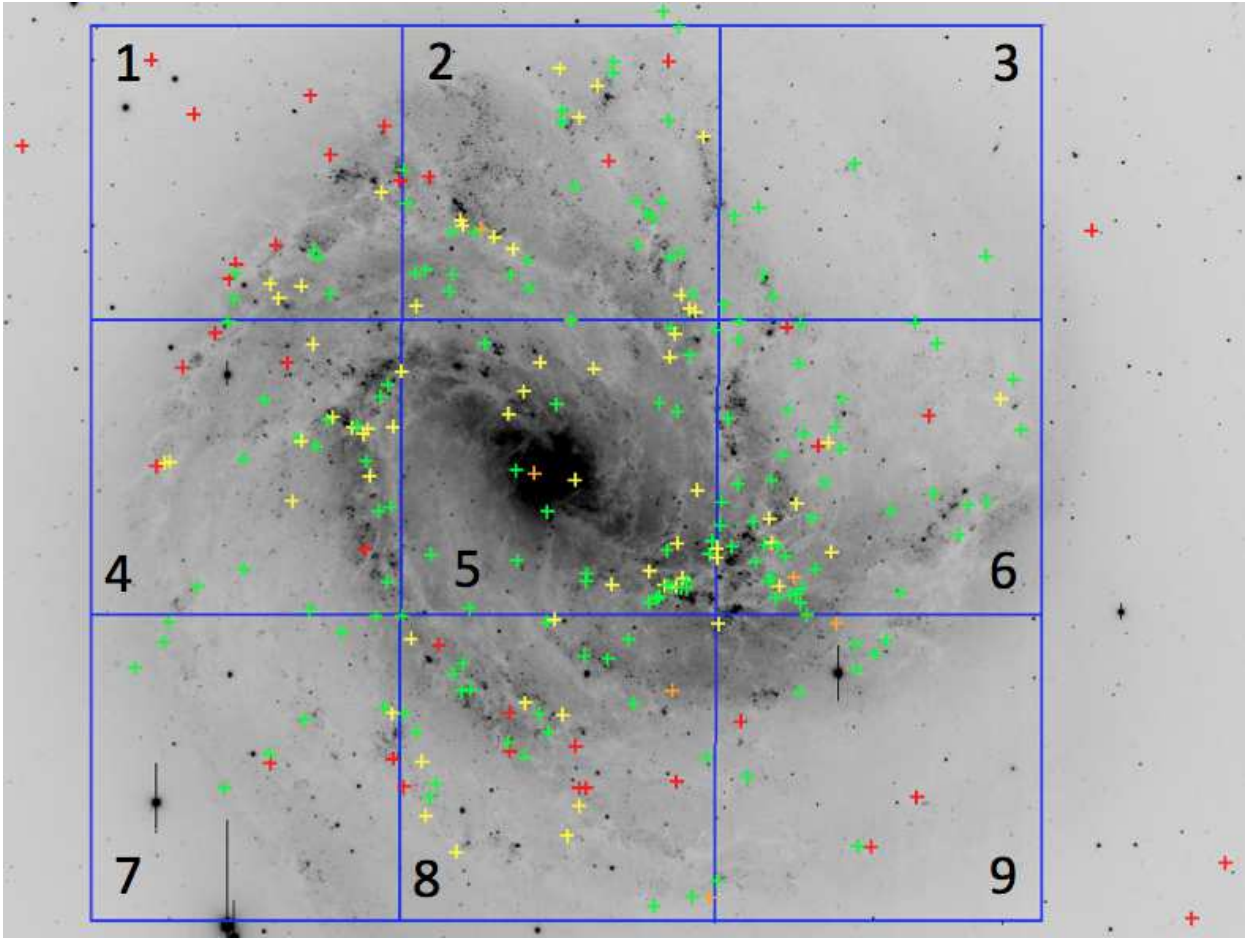}
\caption{The V-band image of M83 from Magellan is shown with a grid of numbered sub-fields that show the approximate locations of the finder charts shown in Figures 5 -- 13.  The colored crosses provide an overview of the spatial distribution of various sources described in the paper: yellow -- normal ISM SNRs with X-ray counterparts; green -- normal ISM SNRs with no X-ray detections; orange -- [O~III]-selected nebulae with X-ray counterparts; and magenta -- [O~III]-selected nebulae with no X-ray detections.  As with all Figures in this paper, north is up and east is to the left.  See text for further information.   \label{fig_overview}}
\end{figure}

\begin{figure}
\plotone{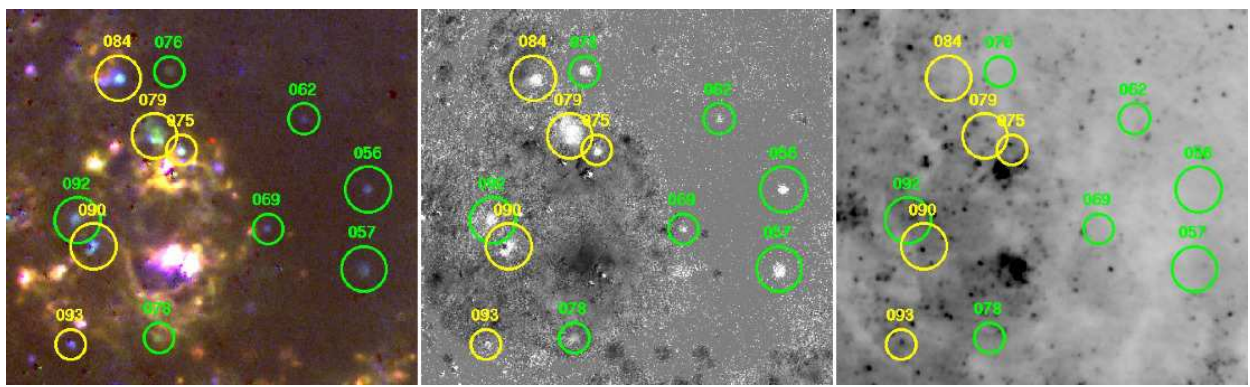}
\caption{A 50\arcsec\ region NW of the nucleus is shown as an example of the search technique for normal SNRs.  The left panel is a three-color representation of the continuum-subtracted H$\alpha$ (red), [S~II] (green), and [O~III] (blue), scaled to maximize the color differences.  Normal SNRs with little or no [O~III] emission will show as green-to-yellow, while SNRs with [O~III] emission show as milky white-to-blue, depending on the strength of [O~III], and  \hii\ regions show as pink-to-orange.  The middle panel is a \sii:H$\alpha$ ratio image, scaled so that enhanced \sii\ nebulae appear white and photoionized gas appears dark.  The right panel show the green continuum band image of the region to provide insight into possible stellar contamination and/or stellar subtraction residuals in the left and middle panels.  The colored circles indicate identified SNRs and have the same meaning as the color definitions in Fig.\ \ref{fig_overview}, with larger circles denoting BL04 SNRs and smaller circles indicating newly discovered SNRs in this paper.  Note how the identified SNRs are found surrounding but not within the bright H~II region complex shown. \label{fig_example}}
\end{figure}

\begin{figure}
\epsscale{1.0}
\plotone{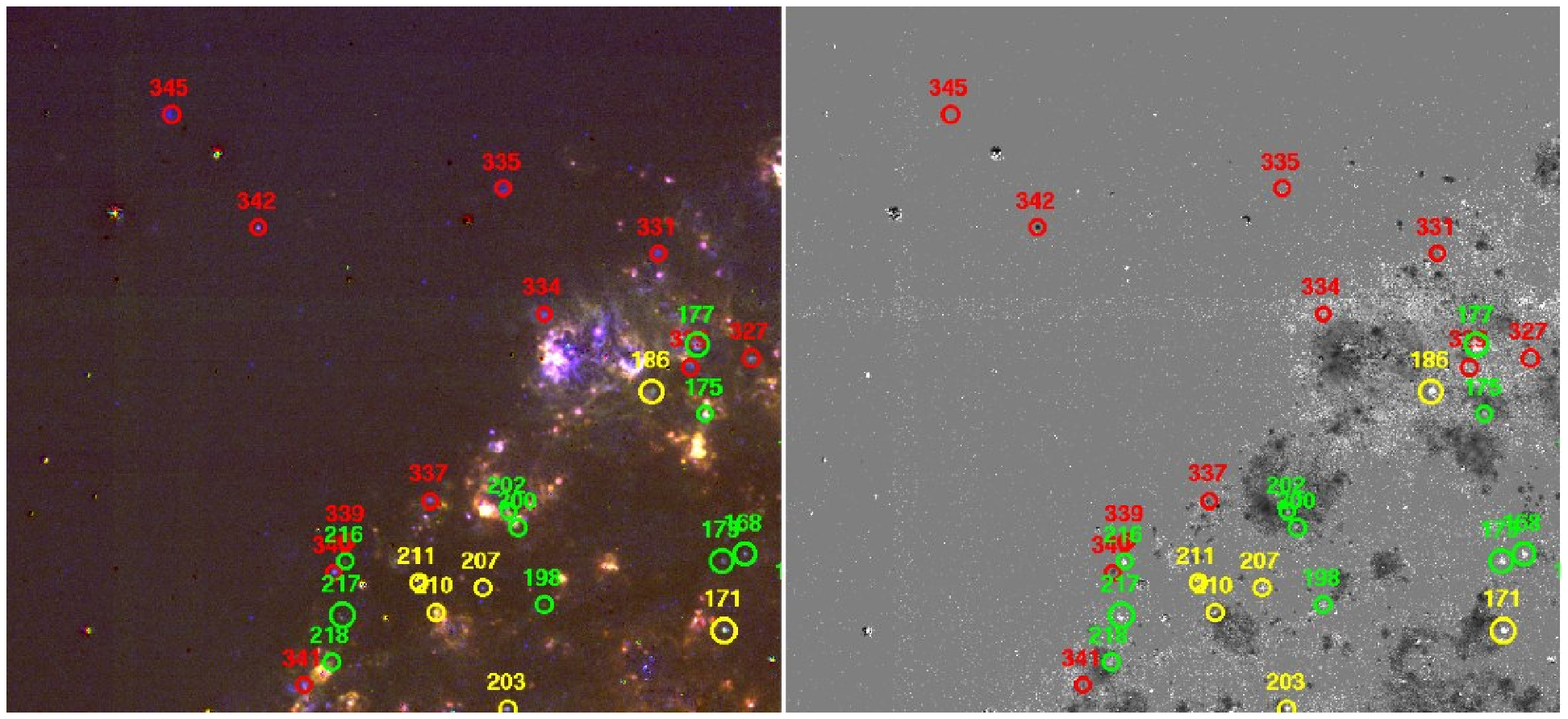}
\caption{A two-panel figure showing the $\sim$3.2\arcmin\ region of Field 1, as indicated in Fig.\ \ref{fig_overview}.  The left panel is a 3 color composite of the continuum-subtracted emission line images, where red
is H$\alpha$, green is [S~II], and blue is [O~III].  The right panel is the [S~II]:H$\alpha$ ratio image of the  same region.  SNR candidates are indicated by the colored circles, as described in the earlier figures, and the ID numbers cross reference to Tables \ref{table_s2} and \ref{table_o3}. \label{fig_F1}}
\end{figure}

\begin{figure}
\plotone{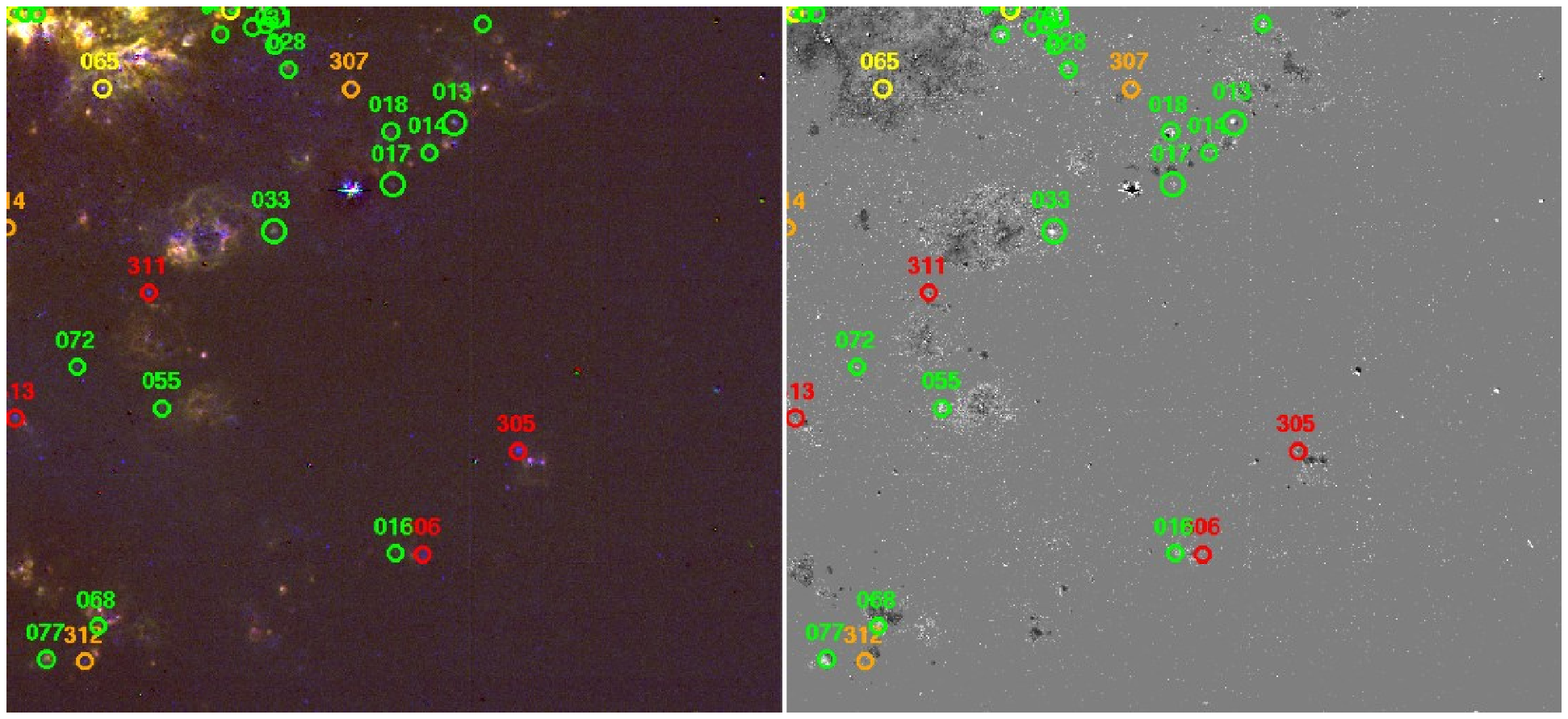}
\caption{Same as Fig.\ \ref{fig_F1} but for Field 9, as indicated in Fig.\ \ref{fig_overview}. \label{fig_F9}}
\end{figure}

Rather than showing individual finder charts for so many objects, we instead display nine $\sim$3.2\arcmin\ regions corresponding approximately to the grid shown in Fig.\   \ref{fig_overview} with all of the SNR candidates marked.  (Each field is slightly larger than the grid shown in Fig.\   \ref{fig_overview} to provide overlap.) These regions are shown in Fig.\  \ref{fig_F1} through Fig.\  \ref {fig_F9}, where yellow circles indicate ISM-dominated SNRs with X-ray counterparts, and green circles  ISM-dominated SNRs with no detected X-ray counterparts.  Likewise, the \oiii-selected objects are shown, with magenta circles indicating the objects with no X-ray counterparts and orange circles indicating the \oiii-selected objects with X-ray emission coincident.  Larger circles are used to identify (previously known) BL04 SNRs and smaller circles indicate our new candidates.  The running identification numbers from Tables \ref{table_s2} and \ref{table_o3} are used to identify the objects in these Figures. 

Referring back now to Fig. \ref{fig_overview}, the global distribution of the SNR population is certainly concentrated at a significant level in the spiral arms and in and around regions of active star formation, perhaps most obvious in sub-field 2 and on the western side of sub-field 5 and on downward into sub-field 8. However, there are clearly a number of SNRs found farther away from the arms and in the interarm regions as well.  Interestingly, with a lone exception on the eastern edge of the nucleus, the \oiii-selected objects we have found are all in the outer part of the galaxy, avoiding sub-field 5 centered on the nucleus.  It is not clear why this is, except that confusion effects from the galaxy background are of course worst in this region.  We also highlight that we have specifically avoided the complex nuclear region where D10 have already reported 20 SNR candidates based on \hst\ data.  Except for a handful of objects on the outer edges of the nucleus that we have also identified, the D10 nuclear SNRs represent additions to those we have tabulated in this work.

\begin{figure}
\epsscale{0.5}
\plotone{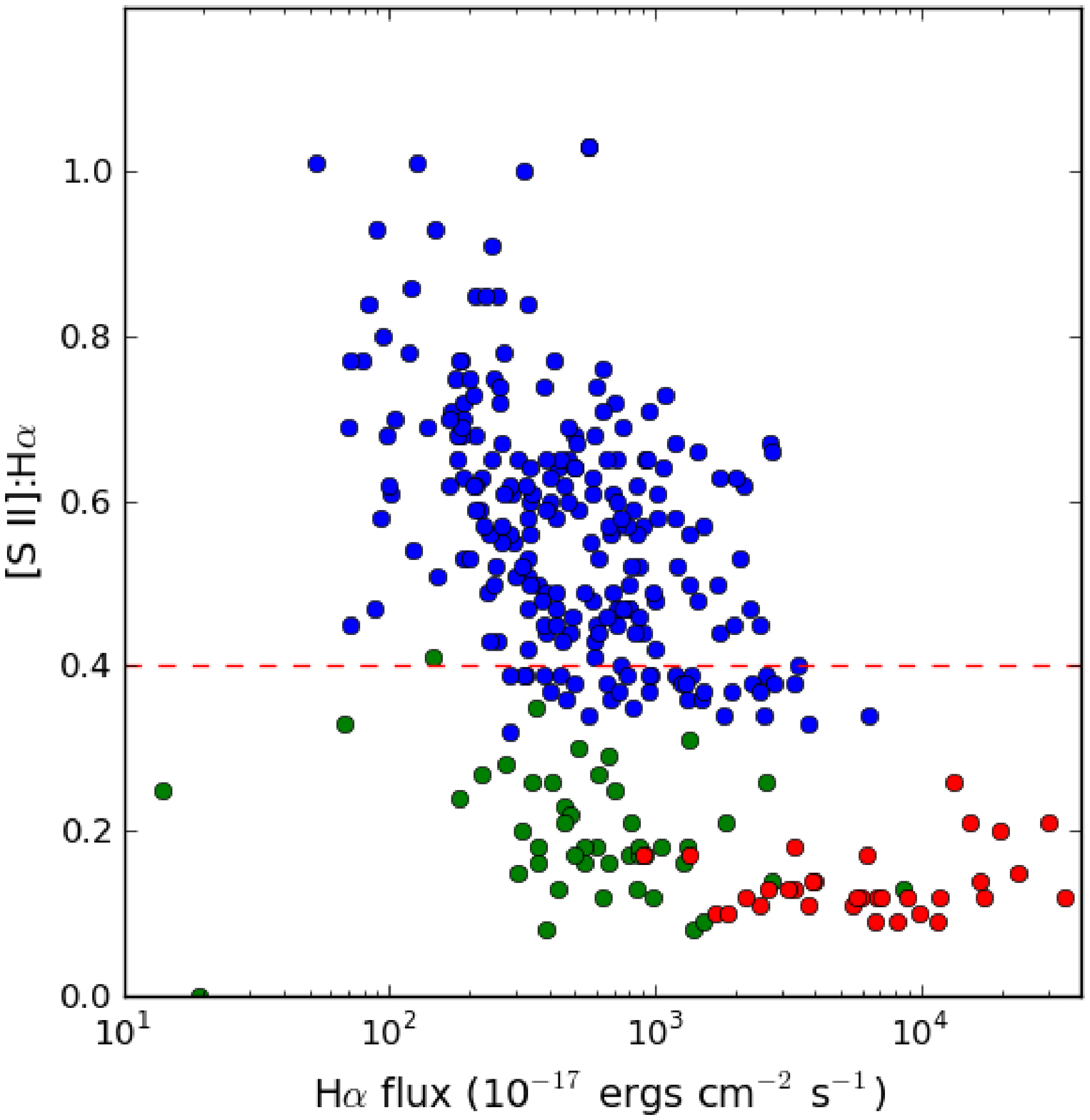}
\plotone{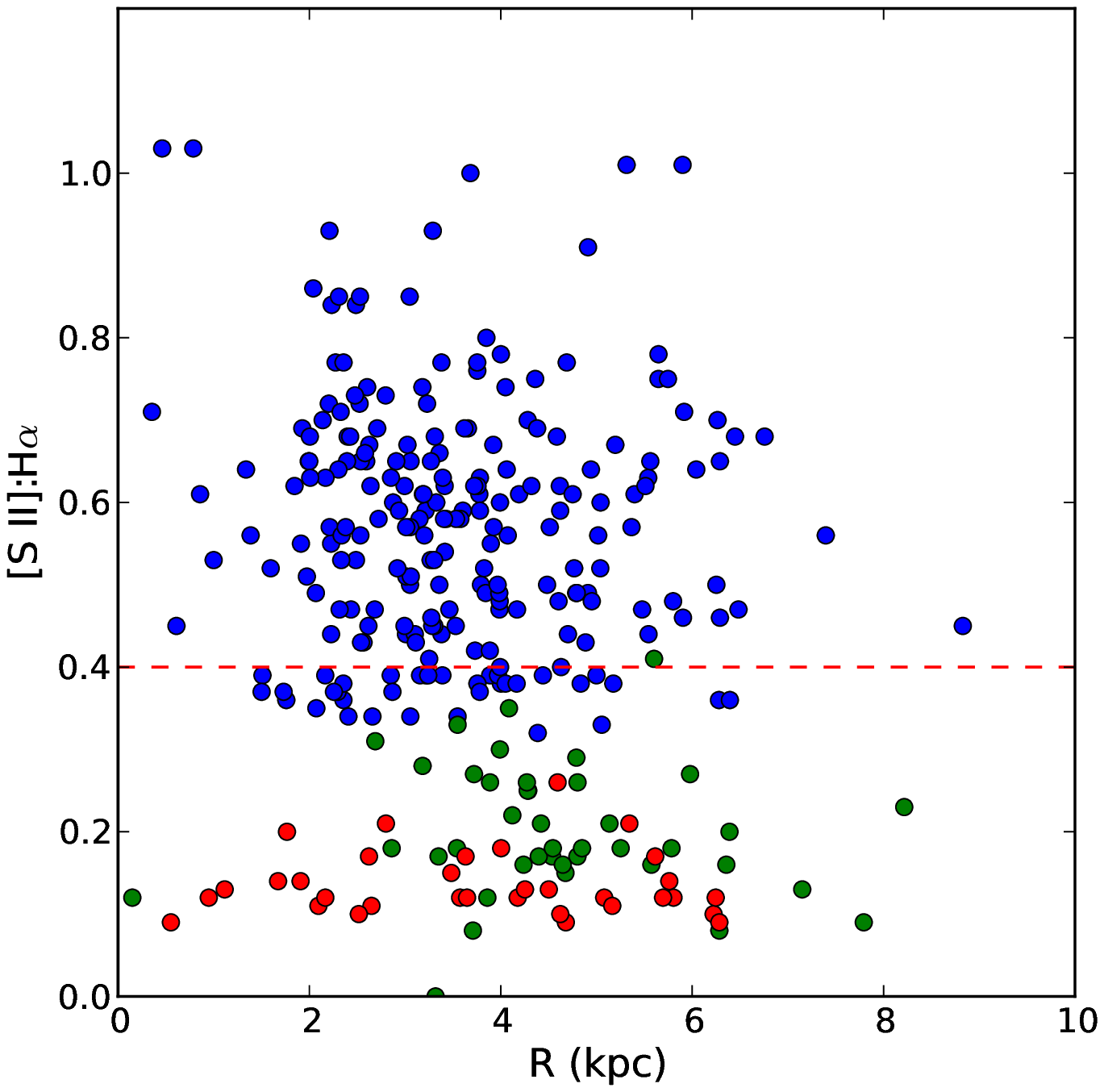}
\caption{Global plots of \sii:\ha\ versus \ha\ flux (top) and galactocentric distance (bottom) for the ISM SNRs (blue), \oiii-selected objects (green) and \hii\ region comparison sample (red; see text).  The \hii\ region sample contains somewhat brighter objects on average, but overlaps with the SNRs and shows no obvious systematic effect in \sii:\ha\ ratio with brightness.  A reference line is drawn at 0.4, and some of the SNR candidates fall somewhat below this line, for reasons described in the text.  There is clear separation in the ratio for ISM SNRs and \hii\ regions, as expected.  Interestingly, the \oiii-selected objects fill in the gap and overlap with both groups, indicating heterogeneity in this group.  None of the groups show an identifiable signature in the \sii:\ha\ ratio as a function of GCD. If present, any such trend could have confused the application of the criterion to find SNRs.\label{fig_global}} 
\end{figure}

As a general check of the selection technique, we have identified a set of 33 \hii\ regions to provide comparison numbers from the imaging analysis.   These objects are relatively compact and isolated, and spatially distributed across the face of the galaxy.  Fluxes and ratios for these objects were extracted in the same manner as described above.  In Fig. \ref{fig_global} we show plots of the image-derived \sii:\ha\ ratio versus F(\ha) and versus the galactocentric distances for all of the objects in Tables \ref{table_s2} and \ref{table_o3} and for the comparison \hii\ regions.  It is clear that there is good separation in the ratio between ISM SNRs and \hii\ regions, with the \oiii-selected objects populating the region between these two groups.  The panel of \sii:\ha\  versus galactocentric distance shows no systematic behavior in \sii:\ha\ for any of the groups of objects, which might have caused confusion.

To quantify these results, for the \hii\ regions, we find an average \sii:\ha\ ratio of 0.14 and $\sigma$=0.05.  For the 225 ISM SNRS in Table \ref{table_s2}, we find an average \sii:\ha\ ratio of 0.57 and $\sigma$=0.16.   From Fig. \ref{fig_global}, it is clear that there are roughly 10 - 15\% of the SNR candidates with image-derived ratios close to or slightly below the nominal 0.4 threshold, with the vast majority well above it.  The objects slightly below the 0.4 threshold  were judged to be candidates in the context of the visual image assessments and fell below 0.4 when the \ha\ flux correction was invoked.   Since any \nii\ contamination of the \ha\ image would tend to decrease the observed ratio, these could still be good SNR candidates and we retain them in the list.  It is clear that in general the \sii\ emission lines in M83 SNRs are very strong, causing a relatively easy separation from photoionized emission regions.

We can do the same experiment for the \oiii:\ha\ ratio.  The \hii\ region sample shows an average \oiii:\ha\ ratio of  0.12 with a modest  $\sigma$=0.13.  For the ISM-dominated SNRs in Table \ref{table_s2}, we find the average \oiii:\ha\ ratio to be 0.49, much higher than for the \hii\ region sample but with a large dispersion of $\sigma$=0.42.  This is expected since the presence or absence of \oiii\ for the SNRs is a function of the shock velocities of the individual objects, which no doubt varies.  The large dispersion is particularly driven by a small number of objects with much higher \oiii:\ha\ values than the majority of objects. Choosing an \oiii:\ha\ ratio $\ge$1.0 to represent unusually high values, there are 21 objects in Table \ref{table_s2} with very strong \oiii\ emission.  Twelve of these objects have sizes $\le$1\arcsec\ ($<$22 pc), and eight of the 21 objects have X-ray detections.  Some of these objects with extreme \oiii:\ha\ ratios and/or with X-ray emission may be young, ejecta-dominated SNRs even though they were found by way of their elevated \sii\ emission.  It should be noted that extinction affects this observed ratio significantly.  In particular, the presence of significant extinction will decrease the observed \oiii:\ha\ ratios.  The sampling of spectroscopic results shown in BL04 shows significant and variable extinction, as one might expect from images of the galaxy that show  prominent dust lanes.  Hence, actual \oiii:\ha\ ratios may by higher than listed. In general, though, this comparison confirms the utility of enhanced \oiii\ emission as a secondary indicator of shock heating for M83 SNRs.

For the \oiii-selected objects in Table \ref{table_o3}, it is not surprising that the average \oiii:\ha\ =1.44 is higher than the average for ISM-dominated SNRs, and much higher than for \hii\ regions. The listed objects have observed \oiii:\ha\ ratios in the range 0.4 to a handful of extreme objects with ratios $\ge$2. They appear to be a somewhat diverse set of objects, as discussed in more detail below, but we believe that many of them are, in fact, SNRs of one kind or another. The second to last column in Table \ref{table_o3} summarizes the discussion below by providing our best estimate of the source ID for each object.

The objects with relatively high ratios of \oiii:\ha\  are not as extreme in ratio as one would expect for some of the well known ejecta-dominated objects like Cas A in our Galaxy  or 1E 0102-7219 in the SMC \citep{blair00}, but some of these objects may be more evolved versions of these young objects that have begun to interact with surrounding material, perhaps more similar to N132D in the LMC \citep{morse96, blair00}, which shows central \oiii\ ejecta knots surrounded by a primary shock front interacting with ISM.  Four of these objects were also identified as O-strong nebulae by D10, and it is interesting that these four objects (307, 309, 327, and 329) also have marginally elevated \sii:\ha\ ratios as well, which is consistent with this interpretation.  Two of these objects, 307 and 309, also have X-ray detections.

Seven of the objects in Table \ref{table_o3} align with X-ray sources, all of which are either unresolved or just barely resolved.  The unresolved sources include the young remnant of SN1957D \citep{long89, long92, milisavljevic12} whose X-ray emission was recently reported by Long et al. (2012a), and the object on the eastern edge of the nuclear region reported previously in the \hst\ data by D10 and corresponding to object 70 in the earlier \citep{soria03} X-ray source catalog. All of the sources with X-ray counterparts are good candidates for young SNRs. For the eight objects with \oiii:\ha\ $\ge$2, the two most extreme (314 and 307) are both X-ray sources, and are strong candidates for new ejecta-dominated SNRs.

\citet{hadfield05} have reported a detailed catalog and spectroscopic follow-up of W-R star candidates in M83.  We have used the catalog information tabulated in their Appendix A Tables A1 (spectrocopically confirmed) and A2 (candidate WR without spectra) to cross check against our \oiii-selected object list, with interesting results.  Six objects from their Table A1 align closely with objects in Table \ref{table_o3}, as shown in the second column from the right, but no additional matches were found from Table A2.  These matches raise the profile of possible WR excitation of other objects in our list since it is likely that the \citet{hadfield05} catalog is far from complete. 

Using the morphology of the matched objects as a guide (e.g. compact but somewhat extended \oiii\ nebulae with little or no \ha\ or \sii), we mark additional objects as possible WR nebulae using the designation `WR?' in Table \ref{table_o3}.  This seems to be especially appropriate for objects at larger galactocentric radii where the mean abundances are lower and the majority of the WR candidates reside.  For instance, objects such as 301, 302, and 303, on the far western side of our survey region, are extended, bright \oiii\ nebulae similar in many respects to objects 345 and 346 on the far northeastern edge, which have WR counterparts in the Hadfield catalog.  For a number of objects, a clear SNR vs. WR designation will require spectroscopic follow-up.  We note in passing that two Hadfield objects align with two normal SNRs (objects 13 and 73 in Table \ref{table_s2}), although in these cases, we conclude the SNR identifications are secure and the WR stars are either chance alignments or are indicative of the youthfulness of the stellar population underlying the SNR positions. \citet{kim12} have also recently noted alignments of Hadfield catalog objects with some of the young stellar clusters in \hst\ WFC3 images. 

%Objects 301, 302, and 303 are all extended well above the PSF of our data and lie at large galactocentric distances (all on the western side of the galaxy).  Hence, whatever their status is, they appear to form a distinct group of nebulae with enhanced \oiii\ emission.  It could be that for some of the sources at galactic radii in excess of about 8 kpc, where the abundances have decreased to sub-solar values and \oiii\ becomes a more effective coolant, we are picking up some normal photoionized \hii\ regions with somewhat elevated ratios, although objects 301, 302, and 303 appear to be too extreme for this explanation.

While most of the objects in Table \ref{table_o3} have low \sii:\ha\ ratios, one of the objects has an observed \sii:\ha\ ratio of 0.41, and several others have ratios above 0.3 which, factoring in \nii\ contamination of the \ha\ data, would in principle qualify them as normal ISM-selected SNR candidates.  A number of these are either X-ray sources or D10 objects already mentioned above. For others, it is possible that they are actually ISM SNRs found by way of their enhanced \oiii\ emission, but for which their \sii\ emission was too low for a good assessment via the \sii:\ha\ ratio image. In many ways, these objects are similar to the strongest \oiii-emitters in the ISM SNR list only fainter.  Object 320 is the BL04 object \#30 moved from the ISM SNR list to the \oiii-selected list.  Its derived \oiii:\ha\ ratio is 2.4, and its \sii:\ha\ ratio is marginally enhanced at 0.27.  The faintness of the \sii\ emission caused this source to appear marginal in the \sii:\ha\ ratio map, again indicating the confusion limit we encounter at the lowest surface brightnesses or in complicated regions of emission.  In general, the \oiii-selected objects fill in the gap in \sii:\ha\ ratio and overlap with both \hii\ regions and ISM SNRs (see again Fig. \ref{fig_global}), pointing to the somewhat heterogeneous nature of this group of objects.

Finally, in assessing the \oiii-selected objects, a few of them align with continuum sources, some of which are also spatially extended, indicating a likely AGN identification.  As with the objects discussed above, spectroscopic follow-up could readily confirm these tentative identifications.

\begin{figure}
\epsscale{1.0}
\plotone{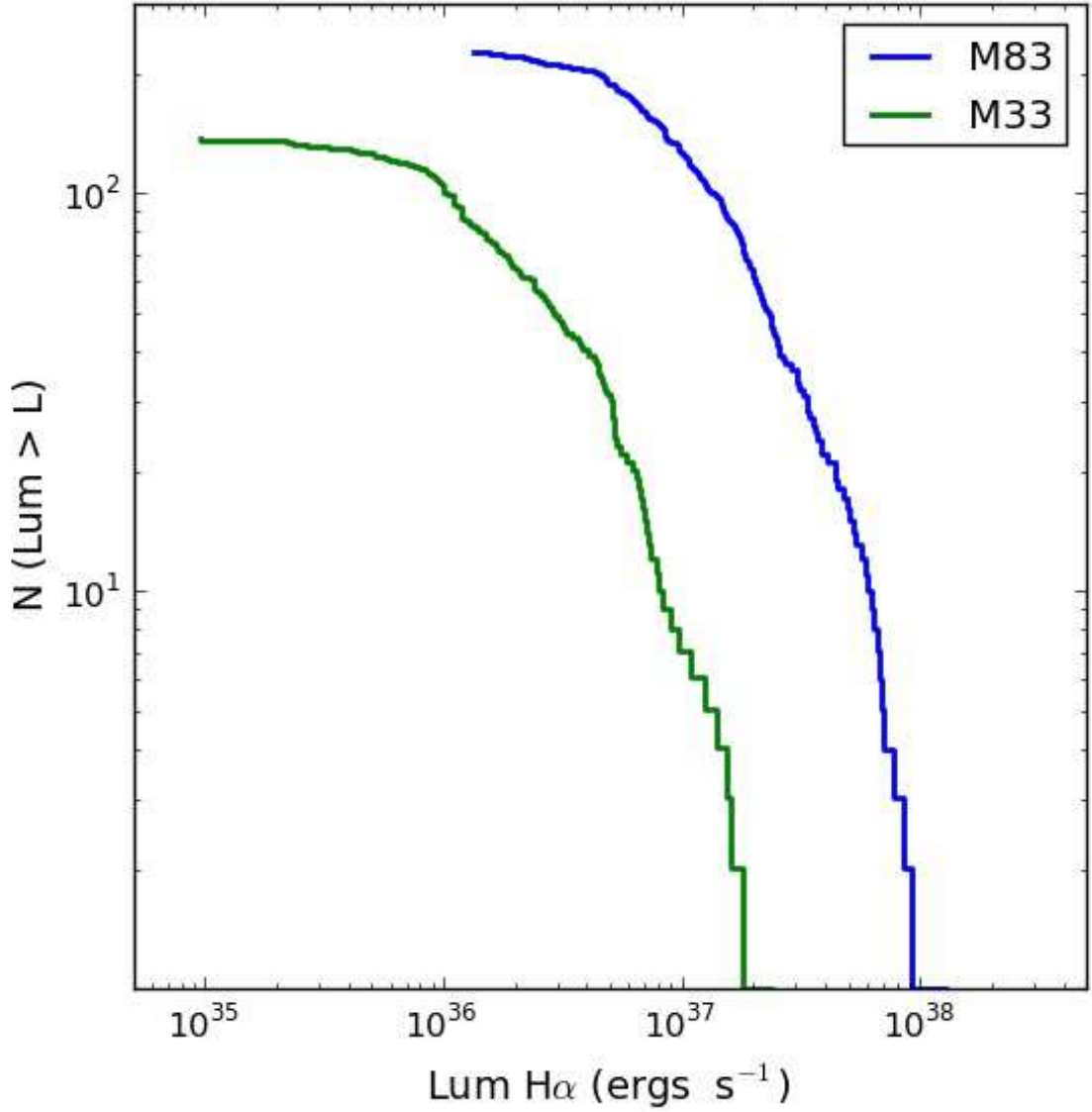}
\caption{The \ha\ Number-luminosity relation for our ISM SNR sample in M83 versus the M33 SNR sample of \citet{long10}. \label{fig_Nlum}}
\end{figure}

In Fig. \ref{fig_Nlum}, we show a cumulative $N(>L_{H\alpha})$ versus $L_{H\alpha}$ plot, comparing the ISM SNR population in M83 to that of SNRs in M33 (Long et al. 2010). The M83 sample is offset toward higher \ha\ luminosities by a factor of $\sim4.5$. This is presumably indicative of higher mean ISM densities in M83 compared with those in M33, since it is unlikely that the SN explosions differ in a major way.  If that is the case, then this density difference would likely be reflected in the density-sensitive ratio of \sii\ $\lambda$6717 to $\lambda$6731, which can ultimately be checked with ground-based spectroscopy.  One would also expect that the more luminous SNRs in M83 would be smaller diameter objects, since high luminosity objects occur in a dense environment, and should reach their peak luminosity faster and fade away more rapidly than objects expanding into less dense material.  At present, any correlation between luminosity and size  is unclear, as objects with luminosities in excess of $\EXPU{1.8}{37}{\LUM}$, the maximum value in M33,  span a large size range in M83.  However, our size estimates are relatively imprecise at present. 

The number of SNRs in M83 exceeds the number  in M33 at all luminosities to which our M83 observations are sensitive, about $\EXPU{1.3}{36}{\LUM}$.  This is surely due to the high rate of star formation in M83.  As a rough estimate, M83's luminosity,  $L_B \approx \EXPU{2.6}{10}{\LSOL}$,   is 4.2 times that of M33 ($\EXPU{6.1}{9}{\LSOL}$).  And there are 70 SNRs in M33 with luminosities exceeding $\EXPU{2}{36}{\LUM}$, compared with   225 in our list of ISM-dominated objects---3.1 times more, roughly consistent with the ratio of $L_B$ for the two galaxies.

While we have expanded the catalog of SNR candidates in M83 dramatically with Magellan/IMACS, it is  clear that the list is still incomplete at a significant level. Even with the excellent seeing conditions, there are limitations to extracting SNRs from regions of bright \hii\ emission where many SNRs are found.  Also, older, low surface brightness  SNRs will systematically be missed in regions of even modest extended emission, compared with low background regions.  Comparing our results in the overlap region of the existing \hst\ data, where we detected 25 of 40 SNRs reported by D10, and scaling to the entire galaxy, another $\sim$100 SNRs could well be detectable in a survey covering all of M83 to the depth of the existing \hst\ observations.

Yet even if the current list were doubled, it is likely that the inventory of SNRs would remain incomplete.   If the six SNe over the past century is typical, and if an average SNR remains visible of 20,000 years, then one would expect some 1200 SNRs to be present.   SNe exploding inside star forming cavities and bubbles will not leave readily detectable remnants at any wavelength, and ones exploding in low density inter-arm regions will likely be relatively faint.  However, the same is true for all spiral galaxies, and M83 clearly stands among the very best venues in which to study large numbers of SNRs and their properties in a systematic way.

\section{Summary \label{sec_summary}}

We have performed a detailed imaging survey of M83 in various emission lines using the 6.5 m Magellan-I telescope and IMACS instrument under conditions of excellent seeing, and we have identified numerous small emission nebulae whose characteristics are consistent with their being supernova remnants. The criterion of strong \sii\ emission relative to \ha\ (ratio $\ge$ 0.4) that has been used in other galaxies to identify shock-heated nebulae works well in M83, and produces a clear  separation between photoionized and shock-heated objects.  We have vetted 68 out of the 71 candidates from our earlier survey (BL04), and have identified an additional 157, for a total of 225 candidates identified by this method.  

Furthermore, the generally low excitation state of most M83 \hii\ regions (due to their high metallicity) has permitted a secondary criterion to be used to search for SNRs.  Small emission nebulae with elevated \oiii:\ha\ emission are expected to be either ejecta-dominated young SNRs, normal SNRs with shock velocities high enough to excite \oiii,  PNe (if point-like) or Wolf-Rayet nebulae (if extended).  Indeed, many of the \sii-selected SNR candidates have significant \oiii\ emission, indicating shock velocities in excess of $\sim$100 $\VEL$ for these objects.  Many point-source \oiii\ emitters have been identified, but we have  identified only seven of these that are also coincident with X-ray sources, raising the likelihood that these objects are young, ejecta-dominated SNRs.  Two of these objects have been identified previously, however, and one corresponds with the young remnant of SN1957D (see Long et  al. 2012a).  We have identified an additional 46 \oiii\ emitters above and beyond the normal SNRs that are either somewhat older ejecta-dominated SNRs akin to N132D in the LMC, normal SNRs with faint \ha\ and \sii\ emission but  whose \oiii\ emission was detectable, or WR nebulae or possibly WR nebulae.  Spectroscopic observations will be required to confirm the actual identifications for these objects. 

Within the single \hst/WFC3 field reported by \citet{dopita10}, we have independently found 25 of their 40 SNR candidates that were outside the complex nuclear region.  The remaining objects were either too faint or too spatially confused in our data to be independently identified.  If this same percentage holds for the extended \hst\ survey that is in progress in \hst\ cycle 19, the total number of M83 SNRs identified through this combination of techniques may well be in excess of 350.  D10 also report 20 SNRs within the complex nuclear region that we have avoided in our search.  The \hst/WFC3 data will supply improved size and morphology information for many of these objects, and future ground-based spectroscopy will be required to solidify the nature of some of the more questionable objects and derive additional physical information about the SNRs reported here.

\acknowledgements

We thank Armin Rest for advice and assistance in the finer points of achieving precise astrometric solutions with the imaging data, Kim Herrmann for sharing her data on M83 to help us resolve the \ha\ flux calibration issue, and Kevin Mather (JHU) for technical assistance.  We also thank the anonymous referee for bringing the WR star connection to our attention for the \oiii-selected nebulae.  WPB acknowledges support from the Dean of the Krieger School of Arts and Sciences and the Center for Astrophysical Sciences at JHU during this work.
PFW and KSL are grateful for Magellan observing support from the staff at Las Campanas Observatory, through time granted via NOAO.
%[PFW and WPB are grateful for both observing and travel support for the Gemini observations from the Gemini office at NOAO.  -- omit if also omitting Gemini data]   
%The Gemini Observatory is operated by Association of Universities for Research in Astronomy, Inc., under a cooperative agreement
%with the NSF on behalf of the Gemini partnership: the National Science Foundation (United
%States), the Science and Technology Facilities Council (United Kingdom), the
%National Research Council (Canada), CONICYT (Chile), the Australian Research Council (Australia),
%Minist\'{e}rio da Ci\^{e}ncia, Tecnologia e Inova\c{c}\~{a}o (Brazil)
%and Ministerio de Ciencia, Tecnologia e Innovaci\'{o}n Productiva (Argentina).
PFW also  acknowledges financial support from the National Science Foundation 
through grant AST-0908566.
Support for this work was also provided by the National Aeronautics and Space Administration 
through \chandra\ Grant Number G01-12115, issued by the \chandra\ X-ray Observatory Center, 
which is operated by the Smithsonian Astrophysical Observatory 
for and on behalf of NASA under contract NAS8-03060\@.  

\bibliographystyle{apj}

%\begin{thebibliography}{26}
\expandafter\ifx\csname natexlab\endcsname\relax\def\natexlab#1{#1}\fi

\clearpage

%\input{tab_obs.tex}

%  SN1006 core sample paper,  Table 1
\begin{deluxetable}{clcrr}
%\tabletypesize{\scriptsize}
%\rotate
%\tablewidth{600pt}
\tablewidth{0pt}
\tablecaption{Magellan/IMACS Observations of M83}

\tablehead{
\colhead {} & \multicolumn{3}{c}{Filter} & 
\colhead {}\\ 

\cline{2-4}  

\colhead{Line} &
\colhead{Designation} &
\colhead{$\rm \lambda_{c}$(\AA)} &
%\colhead{$\rm \lambda_{c}[corr]$(\AA)\tablenotemark{b}} &
\colhead{$\Delta \lambda$(\AA)\tablenotemark{a}} &
\colhead {Exposure (s)} 
}

\startdata

[O III] & CTIO c6014 & 5007 &50\phn\phn & $7\times600$ \phn\phn  \\

Green Cont. & MMTF 5290-156 & 5316& 161\phn\phn&$7\times200$ \phn\phn \\

H$\alpha$ & H$\alpha$ 656 & 6552\tablenotemark{b} & 45\phn\phn & $7\times600$ \phn\phn  \\

[S II] & CTIO c6013 & 6732 & 80\phn\phn & $7\times600$ \phn\phn  \\

Red Cont. & MMTF 7045-228 & 7041& 238\phn\phn&$7\times200$ \phn\phn 

\enddata

%\tablenotetext{a}{Nominal central wavelength.}
\tablenotetext{a}{Full width at half maximum in the Magellan/IMACS f/4.3 beam.}
\tablenotetext{b}{Estimated central wavelength based on comparison of image fluxes with spectra folded through filter curves; value is shifted to blue by 14 \AA\ relative to nominal.}

\label{im_log}
\end{deluxetable}

%\begin{center}
\begin{deluxetable}{cccccrrrcccl}
\tablecaption{ISM-dominated Supernova Remnants and Candidates }
\tablehead{\colhead{Object} & 
 \colhead{RA} & 
 \colhead{Dec.} & 
 \colhead{R} & 
 \colhead{D(ext)$^a$} & 
 \colhead{F(H$\alpha$)$^b$} & 
 \colhead{F([S~II])$^b$} & 
 \colhead{F([O~III])$^b$} & 
 \colhead{[S~II]/H$\alpha$} & 
 \colhead{[O~III]/H$\alpha$} & 
 \colhead{X-ray?$^c$} & 
 \colhead{Other~Names$^d$} 
\\
\colhead{~} & 
 \colhead{(J2000)} & 
 \colhead{(J2000)} & 
 \colhead{(kpc)} & 
 \colhead{(arcsec)} & 
 \colhead{~} & 
 \colhead{~} & 
 \colhead{~} & 
 \colhead{~} & 
 \colhead{~} & 
 \colhead{~} & 
 \colhead{~} 
}
\tabletypesize{\scriptsize}
\tablewidth{0pt}\rotate\startdata
001 &  13:36:39.99 &  -29:51:35.2 &  6.4 &  2.8 &  18.4 &  12.6 &  13.5 &  0.68 &  0.73 &  n &  ~ \\ 
002 &  13:36:40.35 &  -29:51:06.7 &  6.5 &  1.4 &  8.8 &  4.2 &  17.0 &  0.47 &  1.93 &  n &  ~ \\ 
003 &  13:36:40.90 &  -29:51:17.7 &  6.3 &  1.4 &  36.5 &  18.3 &  18.2 &  0.50 &  0.50 &  y &  SW001 \\ 
004 &  13:36:41.50 &  -29:52:16.0 &  5.9 &  2.0 &  5.3 &  5.3 &  11.3 &  1.01 &  2.14 &  n &  ~ \\ 
005 &  13:36:41.58 &  -29:49:56.4 &  6.8 &  1.8 &  9.8 &  6.6 &  6.2 &  0.68 &  0.63 &  n &  ~ \\ 
006 &  13:36:42.33 &  -29:52:17.3 &  5.6 &  2.0 &  11.9 &  9.3 &  1.8 &  0.78 &  0.15 &  n &  ~ \\ 
007 &  13:36:42.73 &  -29:52:34.9 &  5.5 &  1.4 &  19.0 &  12.0 &  11.2 &  0.63 &  0.59 &  n &  ~ \\ 
008 &  13:36:43.70 &  -29:50:45.9 &  5.6 &  2.0 &  17.6 &  13.1 &  4.5 &  0.75 &  0.25 &  n &  ~ \\ 
009 &  13:36:43.83 &  -29:52:11.3 &  5.2 &  2.4 &  26.4 &  17.8 &  5.1 &  0.67 &  0.19 &  n &  BL01 \\ 
010 &  13:36:44.65 &  -29:50:34.2 &  5.5 &  3.2 &  71.6 &  33.9 &  34.5 &  0.47 &  0.48 &  n &  BL02 \\ 
011 &  13:36:45.31 &  -29:53:07.7 &  4.9 &  3.2 &  24.1 &  21.9 &  1.0 &  0.91 &  0.04 &  n &  ~ \\ 
012 &  13:36:45.70 &  -29:52:20.9 &  4.6 &  5.6 &  343.2 &  136.8 &  27.7 &  0.40 &  0.08 &  n &  ~ \\ 
013 &  13:36:45.93 &  -29:53:34.4 &  4.9 &  2.0 &  49.2 &  31.7 &  14.7 &  0.64 &  0.30 &  n &  BL03;H12 \\ 
014 &  13:36:46.42 &  -29:53:42.3 &  4.9 &  2.4 &  25.5 &  11.1 &  13.6 &  0.43 &  0.53 &  n &  ~ \\ 
015 &  13:36:46.93 &  -29:46:41.9 &  8.8 &  4.6 &  60.3 &  26.9 &  6.4 &  0.45 &  0.11 &  n &  ~ \\ 
016 &  13:36:47.13 &  -29:55:31.7 &  6.3 &  4.4 &  48.5 &  22.6 &  8.0 &  0.46 &  0.16 &  n &  ~ \\ 
017 &  13:36:47.18 &  -29:53:51.4 &  4.8 &  2.0 &  23.3 &  11.4 &  4.2 &  0.49 &  0.18 &  n &  BL04 \\ 
018 &  13:36:47.22 &  -29:53:36.9 &  4.6 &  2.4 &  16.8 &  10.4 &  7.8 &  0.62 &  0.47 &  n &  ~ \\ 
019 &  13:36:47.30 &  -29:49:03.9 &  6.0 &  1.6 &  42.8 &  27.3 &  22.1 &  0.64 &  0.52 &  n &  BL05 \\ 
020 &  13:36:47.83 &  -29:51:18.2 &  4.2 &  2.8 &  79.4 &  37.5 &  20.6 &  0.47 &  0.26 &  n &  BL06 \\ 
021 &  13:36:47.91 &  -29:51:45.7 &  4.0 &  3.6 &  331.7 &  109.4 &  10.4 &  0.36 &  0.03 &  n &  ~ \\ 
022 &  13:36:48.10 &  -29:51:33.7 &  4.0 &  2.2 &  33.7 &  20.1 &  3.1 &  0.60 &  0.09 &  n &  ~ \\ 
023 &  13:36:48.31 &  -29:52:44.7 &  3.9 &  2.2 &  269.5 &  179.6 &  176.6 &  0.67 &  0.66 &  y &  BL07;SW007 \\ 
024 &  13:36:48.45 &  -29:51:42.3 &  3.9 &  3.0 &  9.4 &  7.5 &  7.4 &  0.80 &  0.79 &  y &  ~ \\ 
025 &  13:36:48.57 &  -29:52:05.2 &  3.8 &  3.0 &  63.4 &  48.2 &  43.1 &  0.76 &  0.68 &  n &  BL08 \\ 
026 &  13:36:48.99 &  -29:52:54.1 &  3.8 &  1.6 &  10.2 &  6.2 &  3.5 &  0.61 &  0.35 &  n &  ~ \\ 
027 &  13:36:49.12 &  -29:52:24.9 &  3.6 &  2.2 &  51.2 &  30.2 &  7.7 &  0.59 &  0.15 &  n &  ~ \\ 
028 &  13:36:49.37 &  -29:53:20.1 &  3.9 &  3.2 &  119.2 &  46.9 &  5.9 &  0.39 &  0.05 &  n &  ~ \\ 
029 &  13:36:49.51 &  -29:51:37.2 &  3.5 &  2.6 &  55.6 &  17.3 &  7.1 &  0.34 &  0.13 &  n &  ~ \\ 
030 &  13:36:49.63 &  -29:53:05.5 &  3.7 &  4.4 &  31.9 &  31.9 &  7.3 &  1.00 &  0.23 &  n &  BL09 \\ 
031 &  13:36:49.64 &  -29:53:13.5 &  3.8 &  2.8 &  18.6 &  14.4 &  12.7 &  0.77 &  0.68 &  n &  ~ \\ 
032 &  13:36:49.67 &  -29:50:34.7 &  4.1 &  3.0 &  27.6 &  15.4 &  4.2 &  0.56 &  0.15 &  n &  ~ \\ 
033 &  13:36:49.69 &  -29:54:04.2 &  4.4 &  1.8 &  24.7 &  18.5 &  6.8 &  0.75 &  0.28 &  n &  BL10 \\ 
034 &  13:36:49.73 &  -29:50:57.5 &  3.8 &  1.8 &  78.9 &  39.4 &  25.3 &  0.50 &  0.32 &  n &  ~ \\ 
035 &  13:36:49.82 &  -29:53:08.3 &  3.7 &  1.4 &  7.0 &  4.8 &  1.0 &  0.69 &  0.15 &  n &  ~ \\ 
036 &  13:36:49.81 &  -29:52:17.0 &  3.4 &  1.6 &  38.6 &  16.9 &  20.8 &  0.44 &  0.54 &  y &  SW010 \\ 
037 &  13:36:50.13 &  -29:53:08.8 &  3.6 &  1.0 &  9.3 &  5.4 &  1.6 &  0.58 &  0.18 &  n &  ~ \\ 
038 &  13:36:50.22 &  -29:51:24.0 &  3.4 &  1.6 &  12.3 &  6.7 &  63.4 &  0.54 &  0.06 &  n &  ~ \\ 
039 &  13:36:50.29 &  -29:52:47.3 &  3.4 &  2.4 &  275.6 &  181.8 &  63.4 &  0.66 &  0.23 &  n &  BL11 \\ 
040 &  13:36:50.45 &  -29:51:49.3 &  3.2 &  3.2 &  82.9 &  49.0 &  14.3 &  0.59 &  0.17 &  n &  ~ \\ 
041 &  13:36:50.56 &  -29:53:03.9 &  3.4 &  0.8 &  9.9 &  6.1 &  5.0 &  0.62 &  0.51 &  y &  ~ \\ 
042 &  13:36:50.68 &  -29:52:41.4 &  3.2 &  2.0 &  259.4 &  101.9 &  93.7 &  0.39 &  0.36 &  n &  ~ \\ 
043 &  13:36:50.76 &  -29:53:10.6 &  3.4 &  2.6 &  42.3 &  24.4 &  3.7 &  0.58 &  0.09 &  n &  ~ \\ 
044 &  13:36:50.84 &  -29:50:18.9 &  4.0 &  1.6 &  44.2 &  14.6 &  0.3 &  0.36 &  0.01 &  n &  ~ \\ 
045 &  13:36:50.85 &  -29:52:39.6 &  3.2 &  1.8 &  119.1 &  68.6 &  54.5 &  0.58 &  0.46 &  y &  SW011 \\ 
046 &  13:36:50.91 &  -29:52:03.7 &  3.0 &  1.4 &  21.2 &  18.0 &  7.1 &  0.85 &  0.34 &  n &  ~ \\ 
047 &  13:36:50.93 &  -29:52:58.5 &  3.3 &  1.6 &  46.7 &  30.1 &  14.1 &  0.65 &  0.30 &  n &  ~ \\ 
048 &  13:36:51.00 &  -29:52:25.6 &  3.1 &  2.8 &  133.2 &  66.8 &  144.5 &  0.50 &  1.08 &  y &  SW012 \\ 
049 &  13:36:51.02 &  -29:53:01.1 &  3.3 &  1.0 &  19.0 &  10.1 &  9.3 &  0.53 &  0.49 &  n &  ~ \\ 
050 &  13:36:51.25 &  -29:52:40.6 &  3.1 &  2.8 &  639.4 &  195.1 &  80.8 &  0.37 &  0.13 &  n &  ~ \\ 
051 &  13:36:51.33 &  -29:50:07.1 &  4.0 &  1.4 &  65.8 &  25.1 &  2.7 &  0.38 &  0.04 &  n &  ~ \\ 
052 &  13:36:51.47 &  -29:49:28.5 &  4.7 &  2.6 &  8.0 &  6.2 &  12.1 &  0.77 &  1.51 &  n &  ~ \\ 
053 &  13:36:51.60 &  -29:52:50.0 &  3.0 &  1.6 &  33.4 &  17.1 &  2.0 &  0.51 &  0.06 &  n &  ~ \\ 
054 &  13:36:51.70 &  -29:52:27.5 &  2.9 &  2.0 &  96.0 &  37.4 &  12.1 &  0.39 &  0.13 &  n &  ~ \\ 
055 &  13:36:51.98 &  -29:54:52.1 &  4.8 &  3.0 &  28.6 &  17.4 &  4.3 &  0.61 &  0.15 &  n &  ~ \\ 
056 &  13:36:52.34 &  -29:50:33.2 &  3.4 &  1.8 &  22.0 &  13.9 &  13.0 &  0.63 &  0.59 &  n &  BL12 \\ 
057 &  13:36:52.39 &  -29:50:43.7 &  3.2 &  2.4 &  25.8 &  18.6 &  16.9 &  0.72 &  0.65 &  n &  BL13 \\ 
058 &  13:36:52.38 &  -29:52:05.2 &  2.6 &  1.8 &  30.5 &  19.8 &  8.3 &  0.65 &  0.27 &  n &  ~ \\ 
059 &  13:36:52.55 &  -29:49:33.1 &  4.4 &  2.4 &  28.3 &  9.1 &  19.5 &  0.32 &  0.69 &  n &  ~ \\ 
060 &  13:36:52.65 &  -29:52:41.2 &  2.7 &  3.2 &  258.5 &  86.8 &  23.0 &  0.34 &  0.09 &  n &  ~ \\ 
061 &  13:36:52.80 &  -29:51:28.2 &  2.6 &  2.2 &  71.1 &  32.1 &  9.9 &  0.45 &  0.14 &  n &  ~ \\ 
062 &  13:36:53.00 &  -29:50:23.9 &  3.4 &  1.8 &  7.1 &  5.5 &  9.8 &  0.77 &  1.37 &  n &  ~ \\ 
063 &  13:36:53.07 &  -29:52:16.1 &  2.4 &  4.8 &  49.7 &  34.0 &  42.7 &  0.68 &  0.86 &  n &  ~ \\ 
064 &  13:36:53.18 &  -29:52:29.2 &  2.4 &  2.8 &  79.9 &  37.5 &  3.4 &  0.47 &  0.04 &  n &  ~ \\ 
065 &  13:36:53.24 &  -29:53:25.3 &  3.0 &  1.2 &  48.1 &  21.2 &  26.9 &  0.44 &  0.56 &  y &  SW017 \\ 
066 &  13:36:53.30 &  -29:52:42.5 &  2.5 &  1.2 &  32.9 &  17.4 &  7.2 &  0.53 &  0.22 &  y &  SW018 \\ 
067 &  13:36:53.29 &  -29:52:48.1 &  2.5 &  1.2 &  43.6 &  28.3 &  30.2 &  0.65 &  0.69 &  y &  SW019 \\ 
068 &  13:36:53.32 &  -29:55:51.4 &  5.8 &  3.2 &  57.6 &  27.7 &  7.7 &  0.48 &  0.13 &  n &  ~ \\ 
069 &  13:36:53.36 &  -29:50:38.4 &  3.1 &  1.6 &  15.1 &  7.7 &  11.3 &  0.51 &  0.75 &  n &  ~ \\ 
070 &  13:36:53.51 &  -29:52:38.1 &  2.4 &  2.6 &  24.0 &  15.7 &  10.3 &  0.65 &  0.43 &  n &  ~ \\ 
071 &  13:36:53.64 &  -29:52:45.9 &  2.4 &  2.0 &  59.5 &  40.4 &  21.1 &  0.68 &  0.36 &  n &  BL14 \\ 
072 &  13:36:53.77 &  -29:54:41.1 &  4.3 &  2.0 &  10.5 &  7.3 &  10.2 &  0.70 &  0.97 &  n &  ~ \\ 
073 &  13:36:53.89 &  -29:48:48.4 &  5.1 &  2.2 &  378.2 &  111.6 &  111.2 &  0.33 &  0.29 &  y &  H49;SW020 \\ 
074 &  13:36:54.15 &  -29:52:09.3 &  2.1 &  1.6 &  39.0 &  19.3 &  30.9 &  0.49 &  0.79 &  y &  BL15;SW022 \\ 
075 &  13:36:54.23 &  -29:50:28.0 &  3.0 &  1.6 &  119.5 &  79.8 &  49.0 &  0.67 &  0.41 &  y &  ~ \\ 
076 &  13:36:54.36 &  -29:50:17.6 &  3.2 &  2.4 &  26.0 &  19.2 &  4.2 &  0.74 &  0.16 &  n &  ~ \\ 
077 &  13:36:54.44 &  -29:56:00.3 &  5.9 &  2.0 &  65.4 &  30.1 &  6.7 &  0.46 &  0.10 &  n &  ~ \\ 
078 &  13:36:54.47 &  -29:50:52.8 &  2.6 &  1.8 &  58.8 &  25.5 &  8.5 &  0.43 &  0.14 &  n &  ~ \\ 
079 &  13:36:54.51 &  -29:50:25.8 &  3.0 &  3.4 &  214.9 &  134.1 &  35.1 &  0.62 &  0.16 &  y &  BL16 \\ 
080 &  13:36:54.61 &  -29:53:04.8 &  2.4 &  2.2 &  181.1 &  60.9 &  0.4 &  0.34 &  0.00 &  n &  ~ \\ 
081 &  13:36:54.62 &  -29:53:01.2 &  2.4 &  2.6 &  132.1 &  41.5 &  8.7 &  0.34 &  0.07 &  n &  ~ \\ 
082 &  13:36:54.78 &  -29:52:59.5 &  2.3 &  2.0 &  94.0 &  29.1 &  5.9 &  0.33 &  0.06 &  y &  ~ \\ 
083 &  13:36:54.85 &  -29:53:04.7 &  2.4 &  1.8 &  125.4 &  48.1 &  14.8 &  0.38 &  0.12 &  n &  ~ \\ 
084 &  13:36:54.86 &  -29:50:18.7 &  3.1 &  1.6 &  72.1 &  47.0 &  65.7 &  0.65 &  0.91 &  y &  BL17 \\ 
085 &  13:36:54.89 &  -29:49:54.1 &  3.5 &  2.4 &  38.4 &  17.3 &  16.3 &  0.45 &  0.43 &  n &  ~ \\ 
086 &  13:36:54.95 &  -29:47:46.0 &  6.3 &  3.6 &  46.2 &  10.9 &  27.2 &  0.33 &  0.59 &  n &  ~ \\ 
087 &  13:36:55.04 &  -29:52:39.5 &  2.0 &  1.6 &  92.2 &  60.0 &  51.8 &  0.65 &  0.56 &  y &  BL18;SW025 \\ 
088 &  13:36:55.03 &  -29:51:24.7 &  2.0 &  2.6 &  17.9 &  11.6 &  15.3 &  0.65 &  0.86 &  n &  BL19 \\ 
089 &  13:36:55.07 &  -29:53:04.5 &  2.3 &  1.2 &  107.4 &  68.8 &  28.5 &  0.64 &  0.27 &  y &  SW024 \\ 
090 &  13:36:55.14 &  -29:50:40.8 &  2.6 &  2.4 &  59.6 &  44.3 &  60.2 &  0.74 &  1.01 &  y &  BL20 \\ 
091 &  13:36:55.22 &  -29:53:05.0 &  2.3 &  1.8 &  41.9 &  32.1 &  17.4 &  0.77 &  0.41 &  n &  ~ \\ 
092 &  13:36:55.30 &  -29:50:37.3 &  2.6 &  2.0 &  50.6 &  34.0 &  21.1 &  0.67 &  0.42 &  n &  BL21 \\ 
093 &  13:36:55.35 &  -29:50:53.9 &  2.3 &  1.4 &  19.9 &  10.5 &  16.4 &  0.53 &  0.82 &  y &  ~ \\ 
094 &  13:36:55.37 &  -29:49:56.7 &  3.4 &  2.8 &  137.0 &  53.3 &  21.2 &  0.39 &  0.15 &  n &  ~ \\ 
095 &  13:36:55.39 &  -29:48:39.2 &  5.0 &  2.2 &  39.9 &  23.8 &  31.6 &  0.60 &  0.79 &  n &  BL22 \\ 
096 &  13:36:55.47 &  -29:53:03.3 &  2.2 &  1.8 &  70.2 &  50.3 &  36.3 &  0.72 &  0.52 &  n &  BL24b \\ 
097 &  13:36:55.48 &  -29:52:43.6 &  1.9 &  1.8 &  74.9 &  51.5 &  13.4 &  0.69 &  0.18 &  n &  BL23 \\ 
098 &  13:36:55.62 &  -29:53:03.5 &  2.2 &  2.0 &  174.3 &  109.9 &  39.8 &  0.63 &  0.23 &  y &  BL24;SW028 \\ 
099 &  13:36:55.66 &  -29:47:37.6 &  6.4 &  3.4 &  68.0 &  19.1 &  4.4 &  0.33 &  0.07 &  n &  ~ \\ 
100 &  13:36:55.73 &  -29:49:25.4 &  4.0 &  3.0 &  75.1 &  35.2 &  16.8 &  0.47 &  0.22 &  n &  ~ \\ 
101 &  13:36:55.80 &  -29:51:19.7 &  1.8 &  2.6 &  28.2 &  17.4 &  30.6 &  0.62 &  1.08 &  n &  BL25 \\ 
102 &  13:36:55.83 &  -29:53:09.1 &  2.2 &  3.0 &  56.7 &  31.1 &  16.7 &  0.55 &  0.30 &  n &  ~ \\ 
103 &  13:36:55.92 &  -29:53:10.9 &  2.2 &  1.2 &  8.4 &  7.1 &  8.1 &  0.84 &  0.96 &  n &  ~ \\ 
104 &  13:36:56.06 &  -29:56:05.7 &  5.9 &  2.6 &  17.2 &  12.3 &  14.7 &  0.71 &  0.85 &  n &  ~ \\ 
105 &  13:36:56.10 &  -29:49:34.9 &  3.7 &  2.4 &  100.1 &  42.0 &  9.8 &  0.42 &  0.10 &  n &  ~ \\ 
106 &  13:36:56.23 &  -29:52:55.2 &  1.9 &  1.2 &  29.6 &  16.1 &  2.9 &  0.55 &  0.10 &  y &  SW029 \\ 
107 &  13:36:56.29 &  -29:53:13.6 &  2.2 &  2.0 &  8.9 &  8.3 &  5.2 &  0.93 &  0.58 &  n &  ~ \\ 
108 &  13:36:56.38 &  -29:49:31.9 &  3.8 &  3.0 &  278.3 &  92.3 &  23.0 &  0.35 &  0.08 &  n &  ~ \\ 
109 &  13:36:56.80 &  -29:49:49.8 &  3.3 &  1.0 &  21.1 &  14.3 &  18.3 &  0.68 &  0.87 &  n &  ~ \\ 
110 &  13:36:56.84 &  -29:49:25.3 &  3.8 &  3.2 &  42.4 &  20.9 &  29.3 &  0.49 &  0.69 &  n &  ~ \\ 
111 &  13:36:56.94 &  -29:54:10.7 &  3.3 &  2.8 &  14.9 &  13.9 &  1.9 &  0.93 &  0.13 &  n &  ~ \\ 
112 &  13:36:57.15 &  -29:53:34.0 &  2.5 &  2.6 &  33.4 &  28.0 &  21.9 &  0.84 &  0.66 &  n &  BL28 \\ 
113 &  13:36:57.86 &  -29:48:06.0 &  5.5 &  2.0 &  89.8 &  39.8 &  10.9 &  0.44 &  0.12 &  n &  ~ \\ 
114 &  13:36:57.88 &  -29:48:12.2 &  5.4 &  3.2 &  26.9 &  16.3 &  0.0 &  0.61 &  0.00 &  n &  ~ \\ 
115 &  13:36:57.88 &  -29:53:02.7 &  1.8 &  1.6 &  149.1 &  47.1 &  96.7 &  0.34 &  0.65 &  y &  SW035 \\ 
116 &  13:36:58.07 &  -29:53:45.1 &  2.6 &  2.0 &  45.6 &  28.3 &  10.5 &  0.62 &  0.23 &  n &  ~ \\ 
117 &  13:36:58.55 &  -29:48:19.7 &  5.2 &  2.0 &  232.5 &  88.3 &  107.6 &  0.38 &  0.46 &  y &  ~ \\ 
118 &  13:36:58.71 &  -29:51:00.5 &  1.5 &  1.2 &  37.8 &  14.9 &  30.0 &  0.39 &  0.79 &  y &  SW041;D10-T2-01 \\ 
119 &  13:36:59.00 &  -29:52:56.6 &  1.5 &  1.2 &  40.1 &  14.7 &  0.0 &  0.37 &  0.00 &  n &  ~ \\ 
120 &  13:36:59.00 &  -29:53:01.3 &  1.6 &  1.4 &  87.3 &  45.7 &  26.7 &  0.52 &  0.31 &  n &  ~ \\ 
121 &  13:36:59.11 &  -29:53:43.5 &  2.5 &  2.2 &  44.6 &  19.3 &  5.0 &  0.43 &  0.11 &  n &  SW042 \\ 
122 &  13:36:59.33 &  -29:55:08.9 &  4.5 &  2.2 &  90.4 &  51.7 &  151.1 &  0.57 &  1.67 &  y &  BL29;SW043 \\ 
123 &  13:36:59.35 &  -29:48:37.8 &  4.7 &  1.6 &  174.5 &  76.9 &  8.6 &  0.44 &  0.05 &  y &  ~ \\ 
124 &  13:36:59.50 &  -29:52:03.7 &  0.5 &  1.4 &  55.7 &  57.2 &  70.6 &  1.03 &  1.27 &  y &  BL31;SW045 \\ 
125 &  13:36:59.50 &  -29:49:16.9 &  3.8 &  2.6 &  40.3 &  25.4 &  9.7 &  0.63 &  0.24 &  n &  BL32;D10-T2-04 \\ 
126 &  13:36:59.67 &  -29:50:32.9 &  2.0 &  2.4 &  18.5 &  12.6 &  6.3 &  0.68 &  0.34 &  n &  D10-T2-05 \\ 
127 &  13:36:59.85 &  -29:55:25.9 &  4.9 &  1.2 &  69.6 &  34.2 &  82.0 &  0.49 &  1.18 &  y &  SW048 \\ 
128 &  13:37:00.03 &  -29:48:33.5 &  4.8 &  2.4 &  25.1 &  13.1 &  4.4 &  0.52 &  0.17 &  n &  ~ \\ 
129 &  13:37:00.04 &  -29:54:17.1 &  3.3 &  1.2 &  7.1 &  3.2 &  2.3 &  0.45 &  0.32 &  y &  SW051 \\ 
130 &  13:37:00.09 &  -29:48:40.3 &  4.6 &  1.4 &  37.4 &  17.8 &  8.5 &  0.48 &  0.23 &  n &  ~ \\ 
131 &  13:37:00.19 &  -29:48:10.0 &  5.3 &  2.6 &  12.7 &  12.8 &  25.0 &  1.01 &  1.97 &  y &  ~ \\ 
132 &  13:37:00.33 &  -29:51:20.8 &  0.9 &  1.8 &  34.2 &  20.7 &  40.7 &  0.61 &  1.19 &  n &  BL33;D10-T2-07 \\ 
133 &  13:37:00.40 &  -29:53:22.9 &  2.0 &  1.6 &  12.1 &  10.4 &  7.6 &  0.86 &  0.62 &  y &  ~ \\ 
134 &  13:37:00.66 &  -29:54:26.6 &  3.5 &  2.0 &  101.3 &  59.1 &  21.7 &  0.58 &  0.21 &  n &  ~ \\ 
135 &  13:37:00.70 &  -29:52:21.7 &  0.6 &  2.2 &  247.9 &  110.8 &  9.7 &  0.45 &  0.04 &  n &  D10-T2-08 \\ 
136 &  13:37:00.75 &  -29:53:23.9 &  2.1 &  2.6 &  83.0 &  29.2 &  9.1 &  0.35 &  0.11 &  n &  ~ \\ 
137 &  13:37:01.02 &  -29:50:56.3 &  1.4 &  1.4 &  28.2 &  15.9 &  34.2 &  0.56 &  1.21 &  y &  SW068;D10-T2-09 \\ 
138 &  13:37:01.06 &  -29:54:15.9 &  3.3 &  2.8 &  61.0 &  32.6 &  31.8 &  0.53 &  0.52 &  n &  ~ \\ 
139 &  13:37:01.16 &  -29:57:10.7 &  7.4 &  2.4 &  33.9 &  19.1 &  25.6 &  0.56 &  0.75 &  n &  BL34 \\ 
140 &  13:37:01.52 &  -29:50:14.7 &  2.4 &  1.8 &  18.5 &  14.2 &  7.6 &  0.77 &  0.41 &  n &  BL36 \\ 
141 &  13:37:01.57 &  -29:49:58.8 &  2.7 &  1.6 &  33.3 &  19.3 &  36.1 &  0.58 &  1.08 &  n &  ~ \\ 
142 &  13:37:01.67 &  -29:54:10.3 &  3.2 &  2.0 &  68.7 &  41.9 &  58.9 &  0.61 &  0.86 &  y &  BL35;SW076 \\ 
143 &  13:37:01.72 &  -29:51:13.3 &  1.0 &  2.4 &  209.5 &  111.9 &  164.2 &  0.53 &  0.78 &  y &  BL37;SW077;D10-T2-12 \\ 
144 &  13:37:01.72 &  -29:54:40.4 &  3.9 &  1.6 &  26.3 &  14.3 &  19.9 &  0.55 &  0.76 &  y &  BL38 \\ 
145 &  13:37:02.04 &  -29:52:49.5 &  1.3 &  3.4 &  49.6 &  31.7 &  19.2 &  0.64 &  0.39 &  n &  BL39;D10-T2-13 \\ 
146 &  13:37:02.09 &  -29:51:58.5 &  0.4 &  2.8 &  94.4 &  66.6 &  3.9 &  0.71 &  0.04 &  n &  D10-T2-14 \\ 
147 &  13:37:02.21 &  -29:49:52.4 &  2.9 &  1.2 &  71.5 &  42.6 &  23.1 &  0.60 &  0.32 &  y &  BL40;SW080 \\ 
148 &  13:37:02.32 &  -29:50:07.0 &  2.5 &  2.4 &  23.7 &  13.3 &  19.3 &  0.56 &  0.82 &  n &  ~ \\ 
149 &  13:37:02.42 &  -29:54:33.0 &  3.8 &  1.8 &  21.0 &  13.0 &  13.4 &  0.62 &  0.64 &  n &  ~ \\ 
150 &  13:37:02.42 &  -29:51:25.7 &  0.8 &  1.4 &  55.7 &  57.5 &  64.2 &  1.03 &  1.15 &  y &  BL41;SW081;D10-T2-16 \\ 
151 &  13:37:03.02 &  -29:49:45.6 &  3.1 &  1.4 &  151.4 &  86.4 &  21.8 &  0.57 &  0.14 &  y &  SW083 \\ 
152 &  13:37:03.46 &  -29:50:46.4 &  1.7 &  2.4 &  72.6 &  22.7 &  7.1 &  0.36 &  0.10 &  n &  ~ \\ 
153 &  13:37:03.90 &  -29:49:42.9 &  3.2 &  1.6 &  94.6 &  36.7 &  32.6 &  0.39 &  0.34 &  n &  ~ \\ 
154 &  13:37:04.05 &  -29:54:02.3 &  3.2 &  3.4 &  85.6 &  48.2 &  37.8 &  0.56 &  0.44 &  n &  ~ \\ 
155 &  13:37:04.13 &  -29:53:16.5 &  2.2 &  3.0 &  26.5 &  15.2 &  23.7 &  0.57 &  0.89 &  n &  ~ \\ 
156 &  13:37:04.41 &  -29:49:38.7 &  3.3 &  1.8 &  196.4 &  88.1 &  138.1 &  0.45 &  0.70 &  y &  SW089 \\ 
157 &  13:37:04.43 &  -29:53:47.6 &  2.9 &  2.2 &  81.4 &  42.3 &  40.5 &  0.52 &  0.50 &  n &  ~ \\ 
158 &  13:37:04.46 &  -29:54:03.5 &  3.3 &  3.4 &  86.6 &  40.0 &  17.3 &  0.46 &  0.20 &  n &  ~ \\ 
159 &  13:37:04.51 &  -29:49:35.8 &  3.4 &  1.8 &  171.1 &  85.4 &  91.1 &  0.50 &  0.53 &  y &  ~ \\ 
160 &  13:37:04.72 &  -29:55:34.8 &  5.4 &  1.8 &  77.8 &  44.1 &  50.4 &  0.57 &  0.65 &  y &  BL42 \\ 
161 &  13:37:04.81 &  -29:53:53.6 &  3.1 &  3.8 &  61.2 &  26.7 &  6.5 &  0.44 &  0.11 &  n &  BL43 \\ 
162 &  13:37:04.82 &  -29:50:06.9 &  2.7 &  2.4 &  47.1 &  32.3 &  18.9 &  0.69 &  0.40 &  n &  BL44 \\ 
163 &  13:37:04.85 &  -29:49:42.1 &  3.2 &  2.2 &  32.9 &  12.9 &  15.2 &  0.39 &  0.46 &  n &  ~ \\ 
164 &  13:37:04.97 &  -29:50:16.3 &  2.5 &  2.2 &  19.1 &  13.8 &  8.0 &  0.72 &  0.42 &  n &  ~ \\ 
165 &  13:37:05.59 &  -29:54:56.3 &  4.6 &  2.0 &  18.0 &  12.2 &  20.3 &  0.68 &  1.13 &  n &  ~ \\ 
166 &  13:37:05.79 &  -29:52:46.1 &  2.0 &  2.6 &  29.8 &  15.3 &  18.7 &  0.51 &  0.63 &  n &  D10-T2-18 \\ 
167 &  13:37:05.87 &  -29:55:04.1 &  4.8 &  2.4 &  53.8 &  26.3 &  9.7 &  0.49 &  0.18 &  n &  ~ \\ 
168 &  13:37:06.01 &  -29:50:04.2 &  2.9 &  1.6 &  38.7 &  25.3 &  22.7 &  0.65 &  0.59 &  n &  BL45 \\ 
169 &  13:37:06.03 &  -29:55:14.3 &  5.0 &  1.4 &  120.3 &  62.8 &  65.3 &  0.52 &  0.54 &  y &  BL46;SW095 \\ 
170 &  13:37:06.16 &  -29:54:43.5 &  4.4 &  1.4 &  14.0 &  9.7 &  34.0 &  0.69 &  2.43 &  y &  SW097 \\ 
171 &  13:37:06.44 &  -29:50:24.9 &  2.6 &  2.0 &  144.9 &  95.0 &  35.3 &  0.66 &  0.24 &  y &  BL47;D10-T2-19 \\ 
172 &  13:37:06.44 &  -29:54:27.3 &  4.1 &  2.2 &  33.9 &  21.6 &  38.9 &  0.64 &  1.15 &  n &  BL48 \\ 
173 &  13:37:06.46 &  -29:50:06.1 &  2.9 &  2.4 &  38.6 &  22.7 &  27.2 &  0.59 &  0.71 &  n &  BL49 \\ 
174 &  13:37:06.65 &  -29:53:33.6 &  3.0 &  4.0 &  71.3 &  40.9 &  24.7 &  0.57 &  0.35 &  y &  BL50;SW100 \\ 
175 &  13:37:06.82 &  -29:49:26.3 &  3.8 &  2.0 &  245.4 &  90.7 &  35.4 &  0.37 &  0.14 &  n &  ~ \\ 
176 &  13:37:06.98 &  -29:54:16.6 &  3.9 &  3.8 &  22.4 &  12.8 &  0.0 &  0.57 &  0.00 &  n &  ~ \\ 
177 &  13:37:07.01 &  -29:49:07.7 &  4.2 &  2.8 &  101.6 &  61.9 &  51.9 &  0.61 &  0.51 &  y &  BL51 \\ 
178 &  13:37:07.07 &  -29:53:20.9 &  2.9 &  1.6 &  57.6 &  36.2 &  18.7 &  0.63 &  0.32 &  y &  BL52;SW102 \\ 
179 &  13:37:07.10 &  -29:51:01.5 &  2.2 &  1.0 &  32.0 &  12.5 &  4.8 &  0.39 &  0.15 &  y &  SW104 \\ 
180 &  13:37:07.47 &  -29:51:33.3 &  2.0 &  2.4 &  201.8 &  126.3 &  80.0 &  0.63 &  0.40 &  y &  BL53;SW105;D10-T2-22 \\ 
181 &  13:37:07.51 &  -29:54:16.1 &  4.0 &  2.2 &  26.7 &  20.8 &  9.7 &  0.78 &  0.36 &  n &  BL54 \\ 
182 &  13:37:07.57 &  -29:52:18.9 &  2.1 &  1.8 &  19.1 &  13.4 &  11.6 &  0.70 &  0.61 &  n &  BL55;D10-T2-23 \\ 
183 &  13:37:07.69 &  -29:51:09.9 &  2.2 &  1.0 &  84.1 &  37.2 &  14.8 &  0.44 &  0.18 &  n &  D10-T2-21 \\ 
184 &  13:37:07.71 &  -29:53:01.2 &  2.7 &  1.4 &  42.1 &  19.8 &  15.0 &  0.47 &  0.36 &  n &  ~ \\ 
185 &  13:37:07.81 &  -29:54:12.8 &  4.0 &  4.0 &  99.6 &  47.7 &  10.2 &  0.48 &  0.10 &  n &  BL56 \\ 
186 &  13:37:07.93 &  -29:49:20.0 &  4.0 &  2.0 &  38.4 &  28.2 &  15.6 &  0.74 &  0.41 &  y &  BL57 \\ 
187 &  13:37:07.99 &  -29:51:16.2 &  2.3 &  1.6 &  195.0 &  63.1 &  15.2 &  0.35 &  0.08 &  n &  ~ \\ 
188 &  13:37:08.09 &  -29:52:21.1 &  2.3 &  2.6 &  25.5 &  21.8 &  10.9 &  0.85 &  0.43 &  n &  D10-T2-25 \\ 
189 &  13:37:08.21 &  -29:53:20.5 &  3.1 &  2.4 &  23.9 &  10.2 &  2.9 &  0.43 &  0.12 &  n &  ~ \\ 
190 &  13:37:08.48 &  -29:52:02.0 &  2.3 &  2.2 &  228.1 &  107.2 &  22.2 &  0.47 &  0.10 &  y &  D10-T2-27 \\ 
191 &  13:37:08.57 &  -29:51:35.0 &  2.3 &  1.4 &  63.2 &  45.0 &  33.8 &  0.71 &  0.54 &  y &  BL58;SW109;D10-T2-28 \\ 
192 &  13:37:08.66 &  -29:51:53.5 &  2.3 &  1.8 &  68.4 &  38.6 &  10.1 &  0.56 &  0.15 &  n &  D10-T2-30 \\ 
193 &  13:37:08.75 &  -29:51:37.5 &  2.4 &  2.0 &  67.2 &  38.1 &  16.3 &  0.57 &  0.24 &  y &  BL59;D10-T2-32 \\ 
194 &  13:37:09.04 &  -29:51:33.3 &  2.5 &  1.2 &  20.6 &  15.0 &  2.9 &  0.73 &  0.14 &  n &  D10-T2-33 \\ 
195 &  13:37:09.22 &  -29:51:33.6 &  2.5 &  1.8 &  23.0 &  19.5 &  10.4 &  0.85 &  0.45 &  y &  D10-T2-34 \\ 
196 &  13:37:09.69 &  -29:53:30.3 &  3.6 &  2.8 &  18.8 &  13.0 &  17.3 &  0.69 &  0.92 &  n &  ~ \\ 
197 &  13:37:10.07 &  -29:51:28.0 &  2.8 &  1.2 &  108.2 &  79.1 &  28.1 &  0.73 &  0.26 &  y &  D10-T2-36 \\ 
198 &  13:37:10.19 &  -29:50:18.1 &  3.5 &  2.0 &  33.1 &  15.6 &  1.8 &  0.47 &  0.05 &  n &  ~ \\ 
199 &  13:37:10.33 &  -29:51:28.8 &  2.9 &  1.6 &  151.3 &  47.8 &  45.8 &  0.35 &  0.30 &  n &  D10-T2-37 \\ 
200 &  13:37:10.74 &  -29:49:57.2 &  3.9 &  1.8 &  33.5 &  14.2 &  5.4 &  0.42 &  0.16 &  n &  ~ \\ 
201 &  13:37:10.78 &  -29:51:44.8 &  3.0 &  2.6 &  42.5 &  19.0 &  41.1 &  0.45 &  0.97 &  n &  BL60;D10-T2-39 \\ 
202 &  13:37:10.94 &  -29:49:52.9 &  4.0 &  1.6 &  74.6 &  30.0 &  6.8 &  0.40 &  0.09 &  n &  ~ \\ 
203 &  13:37:10.96 &  -29:50:46.4 &  3.3 &  1.6 &  47.0 &  28.1 &  25.9 &  0.60 &  0.55 &  y &  ~ \\ 
204 &  13:37:11.09 &  -29:53:17.2 &  3.8 &  2.8 &  21.7 &  12.9 &  37.2 &  0.59 &  1.71 &  n &  ~ \\ 
205 &  13:37:11.34 &  -29:54:19.7 &  4.8 &  3.8 &  130.3 &  50.0 &  13.6 &  0.38 &  0.10 &  n &  ~ \\ 
206 &  13:37:11.47 &  -29:51:41.3 &  3.2 &  1.4 &  58.5 &  35.6 &  36.7 &  0.61 &  0.63 &  y &  BL61 \\ 
207 &  13:37:11.48 &  -29:50:13.4 &  3.8 &  1.6 &  31.5 &  16.5 &  21.4 &  0.52 &  0.68 &  y &  ~ \\ 
208 &  13:37:11.68 &  -29:51:39.4 &  3.3 &  3.8 &  58.7 &  24.1 &  15.5 &  0.41 &  0.26 &  n &  BL62 \\ 
209 &  13:37:11.87 &  -29:52:15.6 &  3.4 &  1.4 &  73.7 &  42.7 &  66.9 &  0.58 &  0.91 &  y &  BL63;SW110 \\ 
210 &  13:37:12.46 &  -29:50:20.3 &  4.0 &  1.4 &  97.5 &  47.8 &  9.8 &  0.49 &  0.10 &  y &  ~ \\ 
211 &  13:37:12.81 &  -29:50:12.2 &  4.2 &  2.6 &  49.9 &  18.7 &  41.5 &  0.38 &  0.83 &  y &  SW115 \\ 
212 &  13:37:12.85 &  -29:54:38.9 &  5.5 &  2.6 &  20.8 &  12.8 &  3.7 &  0.62 &  0.18 &  n &  ~ \\ 
213 &  13:37:13.09 &  -29:51:18.4 &  3.7 &  2.8 &  32.9 &  20.5 &  10.5 &  0.62 &  0.32 &  n &  BL64 \\ 
214 &  13:37:14.01 &  -29:52:54.1 &  4.3 &  2.8 &  84.7 &  52.5 &  31.0 &  0.62 &  0.37 &  n &  ~ \\ 
215 &  13:37:13.97 &  -29:51:51.1 &  4.0 &  2.4 &  24.7 &  12.4 &  29.4 &  0.50 &  1.19 &  n &  BL66 \\ 
216 &  13:37:14.35 &  -29:50:06.4 &  4.6 &  1.8 &  21.0 &  12.4 &  2.6 &  0.59 &  0.13 &  n &  ~ \\ 
217 &  13:37:14.42 &  -29:50:21.3 &  4.5 &  2.4 &  34.0 &  16.9 &  8.3 &  0.50 &  0.24 &  n &  BL67 \\ 
218 &  13:37:14.66 &  -29:50:33.7 &  4.4 &  2.2 &  78.6 &  30.6 &  8.9 &  0.39 &  0.11 &  n &  ~ \\ 
219 &  13:37:14.84 &  -29:54:58.6 &  6.3 &  2.0 &  93.8 &  60.8 &  36.3 &  0.65 &  0.39 &  n &  BL68 \\ 
220 &  13:37:16.03 &  -29:53:04.0 &  5.0 &  2.0 &  28.4 &  10.9 &  25.0 &  0.39 &  0.88 &  n &  ~ \\ 
221 &  13:37:17.21 &  -29:51:53.4 &  5.0 &  1.4 &  142.8 &  68.6 &  93.6 &  0.48 &  0.66 &  y &  SW122 \\ 
222 &  13:37:17.26 &  -29:53:25.0 &  5.6 &  2.2 &  65.6 &  42.5 &  42.8 &  0.65 &  0.65 &  n &  BL69 \\ 
223 &  13:37:17.42 &  -29:51:54.0 &  5.0 &  1.4 &  133.3 &  74.4 &  27.2 &  0.56 &  0.20 &  y &  SW123 \\ 
224 &  13:37:17.49 &  -29:53:35.8 &  5.7 &  2.2 &  20.1 &  15.0 &  4.5 &  0.75 &  0.23 &  n &  BL70 \\ 
225 &  13:37:18.74 &  -29:53:50.6 &  6.3 &  2.4 &  16.7 &  11.8 &  25.3 &  0.70 &  1.51 &  n &  BL71 \\ 
\tablenotetext{a}{Diameter of circular regions used for flux extractions; this is effectively an upper limit to the object sizes.}
\tablenotetext{b}{$\rm 10^{-16} ergs~cm^{-2}s^{-1}$; a correction factor of 1.45 has been applied to H$\alpha$ (see text).}
\tablenotetext{c}{ A ``y'' indicates a likely X-ray detection in deep Chandra data (Long et al. 2012, in prep.).}
\tablenotetext{d}{ BL: Blair \& Long (2004); SW: Soria \& Wu (2003); H: Hadfield et al. (2005); D10: Dopita et al. (2010) Table 2.}
\enddata 
\label{table_s2}
\end{deluxetable}
%\end{center}

%\begin{center}
\begin{deluxetable}{cccccrrrccccl}
\tablecaption{[O~III]-selected Objects and Supernova Remnant Candidates }
\tablehead{\colhead{Object$^a$} & 
 \colhead{RA} & 
 \colhead{Dec.} & 
 \colhead{R} & 
 \colhead{D(ext)$^b$} & 
 \colhead{F(H$\alpha$)$^c$} & 
 \colhead{F([S~II])$^c$} & 
 \colhead{F([O~III])$^c$} & 
 \colhead{[S~II]/H$\alpha$} & 
 \colhead{[O~III]/H$\alpha$} & 
 \colhead{X-ray?$^d$} & 
 \colhead{Source ID$^e$} & 
 \colhead{Other~Names$^f$} 
\\
\colhead{~} & 
 \colhead{(J2000)} & 
 \colhead{(J2000)} & 
 \colhead{(kpc)} & 
 \colhead{(arcsec)} & 
 \colhead{~} & 
 \colhead{~} & 
 \colhead{~} & 
 \colhead{~} & 
 \colhead{~} & 
 \colhead{~} & 
 \colhead{~} & 
 \colhead{~} 
}
\tabletypesize{\scriptsize}
\tablewidth{0pt}\rotate\startdata
301 &  13:36:31.06 &  -29:55:41.0 &  10.1 &  3.2 &  18.2 &  4.4 &  58.0 &  0.24 &  3.18 &  n & WR? &  ~ \\ 
302 &  13:36:32.51 &  -29:56:11.9 &  10.1 &  5.6 &  80.3 &  16.5 &  325.3 &  0.21 &  4.05 &  n & WR? &  ~ \\ 
303 &  13:36:36.94 &  -29:49:41.8 &  8.2 &  3.2 &  45.3 &  10.3 &  114.5 &  0.23 &  2.53 &  n & WR? &  ~ \\ 
304 &  13:36:44.05 &  -29:51:27.1 &  5.3 &  1.4 &  36.3 &  6.7 &  21.7 &  0.18 &  0.60 &  n & SNR?;WR? &  ~ \\ 
305 &  13:36:44.54 &  -29:55:03.5 &  6.4 &  3.2 &  66.8 &  11.0 &  74.2 &  0.16 &  1.11 &  n & SNR? &  ~ \\ 
306 &  13:36:46.56 &  -29:55:31.5 &  6.4 &  3.4 &  31.5 &  6.3 &  26.2 &  0.20 &  0.83 &  n & SNR? &  ~ \\ 
307 &  13:36:47.99 &  -29:53:26.6 &  4.3 &  1.0 &  1.4 &  0.4 &  6.2 &  0.25 &  4.41 &  y & SNR?;AGN? &  SW006;D10-T4-03 \\ 
308 &  13:36:48.89 &  -29:51:44.4 &  3.7 &  3.2 &  38.9 &  3.0 &  51.9 &  0.08 &  1.33 &  n & WR &  H22 \\ 
309 &  13:36:49.91 &  -29:52:59.3 &  3.5 &  1.4 &  6.8 &  2.3 &  10.9 &  0.33 &  1.60 &  y & OSNR &  D10-T4-04 \\ 
310 &  13:36:50.24 &  -29:50:36.9 &  3.9 &  1.6 &  34.5 &  9.1 &  43.1 &  0.26 &  1.25 &  n & SNR? &  ~ \\ 
311 &  13:36:52.27 &  -29:54:20.9 &  4.1 &  2.2 &  47.9 &  10.8 &  46.7 &  0.22 &  0.98 &  n & SNR? &  ~ \\ 
312 &  13:36:53.60 &  -29:56:00.9 &  6.0 &  1.6 &  22.2 &  6.1 &  14.1 &  0.27 &  0.63 &  y & OSNR? &  SW019 \\ 
313 &  13:36:55.06 &  -29:54:54.9 &  4.4 &  2.6 &  45.2 &  9.6 &  29.9 &  0.21 &  0.66 &  n & SNR? &  ~ \\ 
314 &  13:36:55.27 &  -29:54:02.9 &  3.3 &  1.4 &  1.9 &  0.0 &  22.6 &  0.00 &  11.74 &  y & OSNR &  SW026 \\ 
315 &  13:36:55.40 &  -29:48:05.9 &  5.8 &  2.4 &  60.2 &  10.9 &  34.0 &  0.18 &  0.56 &  n & WR &  H55 \\ 
316 &  13:36:58.04 &  -29:49:02.0 &  4.2 &  3.0 &  54.3 &  8.6 &  33.7 &  0.16 &  0.62 &  n & SNR? &  ~ \\ 
317 &  13:36:58.49 &  -29:59:24.2 &  10.5 &  4.0 &  862.1 &  108.6 &  1126.4 &  0.13 &  1.31 & n &  WR? &  ~ \\ 
318 &  13:36:59.02 &  -29:54:58.7 &  4.3 &  2.8 &  70.2 &  17.4 &  44.0 &  0.25 &  0.63 &  n & SNR? &  ~ \\ 
319 &  13:36:59.33 &  -29:54:58.6 &  4.3 &  4.4 &  262.5 &  62.1 &  95.4 &  0.24 &  0.37 &  n & SNR? &  ~ \\ 
320 &  13:36:59.44 &  -29:54:34.8 &  3.7 &  2.2 &  61.6 &  16.3 &  148.9 &  0.27 &  2.42 &  n & OSNR? & BL30 \\ 
321 &  13:37:01.27 &  -29:51:59.9 &  0.1 &  1.0 &  97.6 &  7.9 &  36.7 &  0.08 &  0.38 &  y & OSNR &  SW070 \\ 
322 &  13:37:02.35 &  -29:54:37.5 &  3.9 &  1.8 &  62.8 &  7.5 &  48.5 &  0.12 &  0.77 &  n & SNR? &  ~ \\ 
323 &  13:37:02.38 &  -29:54:15.5 &  3.4 &  1.8 &  79.1 &  13.5 &  38.5 &  0.17 &  0.49 &  n & SNR? &  ~ \\ 
324 &  13:37:03.59 &  -29:49:40.8 &  3.2 &  1.4 &  27.3 &  7.7 &  30.6 &  0.28 &  1.12 &  y & OSNR &  SN57D \\ 
325 &  13:37:04.98 &  -29:59:45.8 &  11.2 &  4.0 &  273.7 &  38.1 &  457.8 &  0.14 &  1.67 &  n & AGN? &  ~ \\ 
326 &  13:37:05.47 &  -29:53:37.3 &  2.9 &  1.4 &  54.5 &  9.8 &  46.4 &  0.18 &  0.85 &  n & SNR? &  ~ \\ 
327 &  13:37:05.87 &  -29:49:11.4 &  4.0 &  1.6 &  51.7 &  15.4 &  33.7 &  0.30 &  0.65 &  n & OSNR? &  D10-T4-02 \\ 
328 &  13:37:06.96 &  -29:54:57.7 &  4.8 &  2.4 &  40.6 &  10.6 &  29.9 &  0.26 &  0.74 &  n & SNR? &  ~ \\ 
329 &  13:37:07.15 &  -29:49:13.4 &  4.1 &  2.8 &  35.5 &  12.3 &  42.6 &  0.35 &  1.20 &  n & OSNR? &  D10-T4-01 \\ 
330 &  13:37:07.46 &  -29:54:42.1 &  4.5 &  2.0 &  50.0 &  8.5 &  35.1 &  0.17 &  0.70 &  n & SNR?;WR? &  ~ \\ 
331 &  13:37:07.81 &  -29:48:42.8 &  4.8 &  3.4 &  91.4 &  15.8 &  66.3 &  0.17 &  0.73 &  n & SNR?;WR? &  ~ \\ 
332 &  13:37:08.19 &  -29:59:19.6 &  10.8 &  4.0 &  85.0 &  11.2 &  168.2 &  0.13 &  1.98 &  n & AGN? &  ~ \\ 
333 &  13:37:08.66 &  -29:52:42.9 &  2.7 &  2.4 &  133.9 &  41.4 &  53.8 &  0.31 &  0.40 &  n & SNR? &  ~ \\ 
334 &  13:37:10.19 &  -29:48:59.2 &  4.8 &  2.6 &  66.6 &  19.2 &  47.7 &  0.29 &  0.72 &  n & SNR? &  ~ \\ 
335 &  13:37:11.05 &  -29:48:25.0 &  5.6 &  3.4 &  36.1 &  5.8 &  50.3 &  0.16 &  1.39 &  n & WR &  H114 \\ 
336 &  13:37:12.08 &  -29:50:57.1 &  3.5 &  1.0 &  130.9 &  15.7 &  52.6 &  0.12 &  0.40 &  y & SNR &  ~ \\ 
337 &  13:37:12.57 &  -29:49:49.8 &  4.4 &  2.0 &  86.8 &  14.8 &  51.2 &  0.17 &  0.59 &  n & WR &  H121 \\ 
338 &  13:37:12.82 &  -29:54:44.7 &  5.6 &  2.4 &  14.5 &  6.0 &  18.3 &  0.41 &  1.26 &  n & SNR? &  ~ \\ 
339 &  13:37:14.30 &  -29:50:00.8 &  4.7 &  2.0 &  30.2 &  4.7 &  63.1 &  0.15 &  2.09 &  n & SNR? &  ~ \\ 
340 &  13:37:14.57 &  -29:50:09.4 &  4.6 &  2.2 &  127.0 &  19.7 &  111.5 &  0.16 &  0.88 &  n & SNR? &  ~ \\ 
341 &  13:37:15.19 &  -29:50:40.0 &  4.5 &  2.2 &  104.8 &  19.2 &  93.8 &  0.18 &  0.89 &  n & SNR?;WR? &  ~ \\ 
342 &  13:37:16.15 &  -29:48:35.8 &  6.3 &  1.8 &  139.5 &  11.0 &  87.1 &  0.08 &  0.62 &  n & AGN?;WR? &  ~ \\ 
343 &  13:37:16.66 &  -29:50:59.8 &  4.8 &  2.8 &  86.4 &  15.8 &  63.0 &  0.18 &  0.73 &  n & SNR? &  ~ \\ 
344 &  13:37:17.80 &  -29:51:55.6 &  5.1 &  4.0 &  184.3 &  39.3 &  216.0 &  0.21 &  1.17 &  n & SNR?;WR? &  ~ \\ 
345 &  13:37:17.99 &  -29:48:04.9 &  7.2 &  4.0 &  42.8 &  5.4 &  99.9 &  0.13 &  2.33 &  n & WR &  H130 \\ 
346 &  13:37:23.65 &  -29:48:53.7 &  7.8 &  4.0 &  151.6 &  14.2 &  106.3 &  0.09 &  0.70 &  n & WR &  H132 \\ 
\tablenotetext{a}{Numbering in this table begins at 301 to separate this group from the ISM SNR sample.}
\tablenotetext{b}{Diameter of circular regions used for flux extractions; this is effectively an upper limit to the object sizes.}
\tablenotetext{c}{ $\rm 10^{-16} ergs~cm^{-2}s^{-1}$; a correction factor of 1.45 has been applied to H$\alpha$ (see text).}
\tablenotetext{d}{ A ``y'' indicates a likely X-ray detection in deep Chandra data (Long et al. 2012, in prep.).}
\tablenotetext{e}{SNR=supernova remnant; OSNR=oxygen-dominated SNR; WR=Wolf-Rayet star; AGN=active galaxy nucleus; question marks denote uncertainty in the ID.} 
\tablenotetext{f}{ BL: Blair \& Long (2004); SW: Soria \& Wu (2003); H: Hadfield et al. (2005); D10: Dopita et al. (2010) Table 4.}
\enddata 
\label{table_o3}
\end{deluxetable}
%\end{center}

%\clearpage

\begin{figure}
\plotone{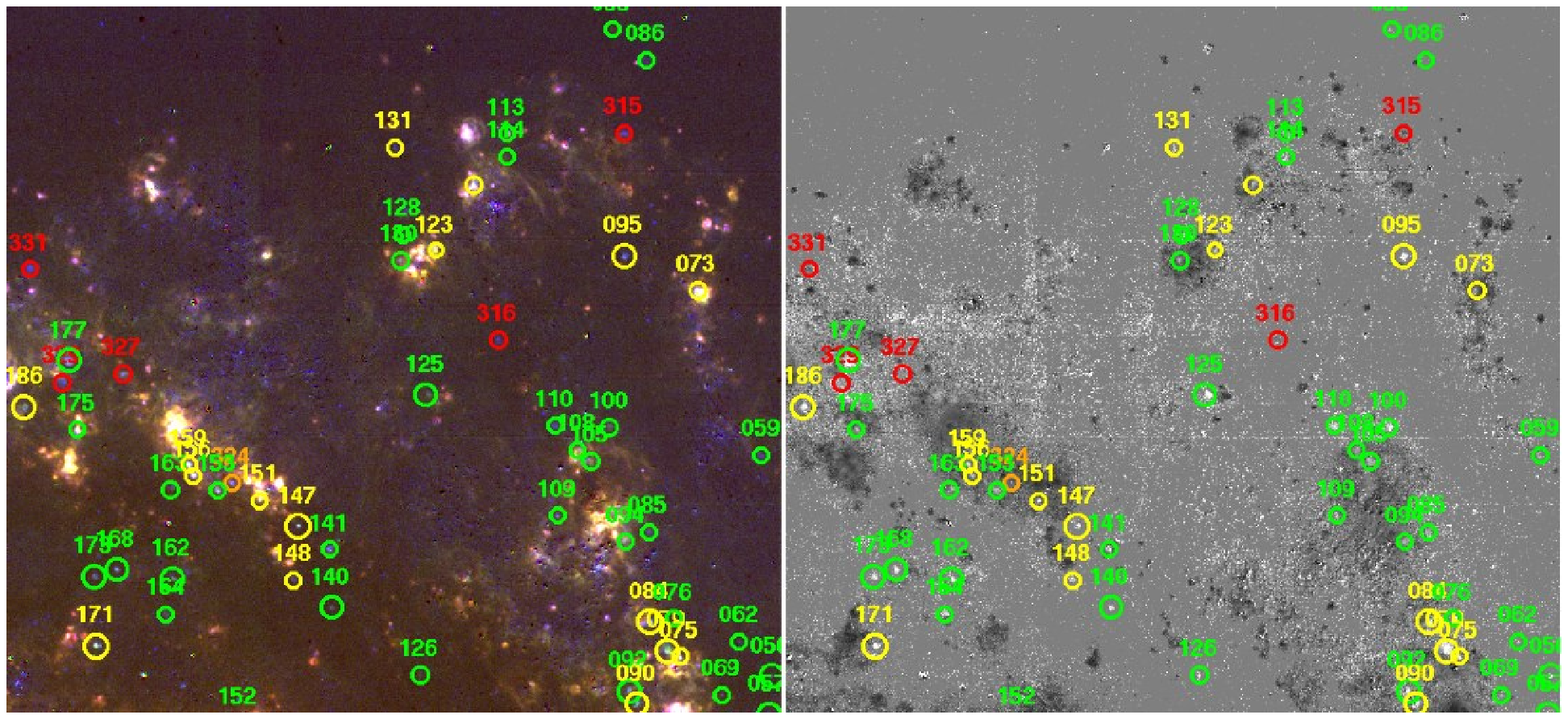}
\caption{Same as Fig.\ \ref{fig_F1} but for Field 2, as indicated in Fig.\ \ref{fig_overview}. \label{fig_F2}}
\end{figure}

\begin{figure}
\plotone{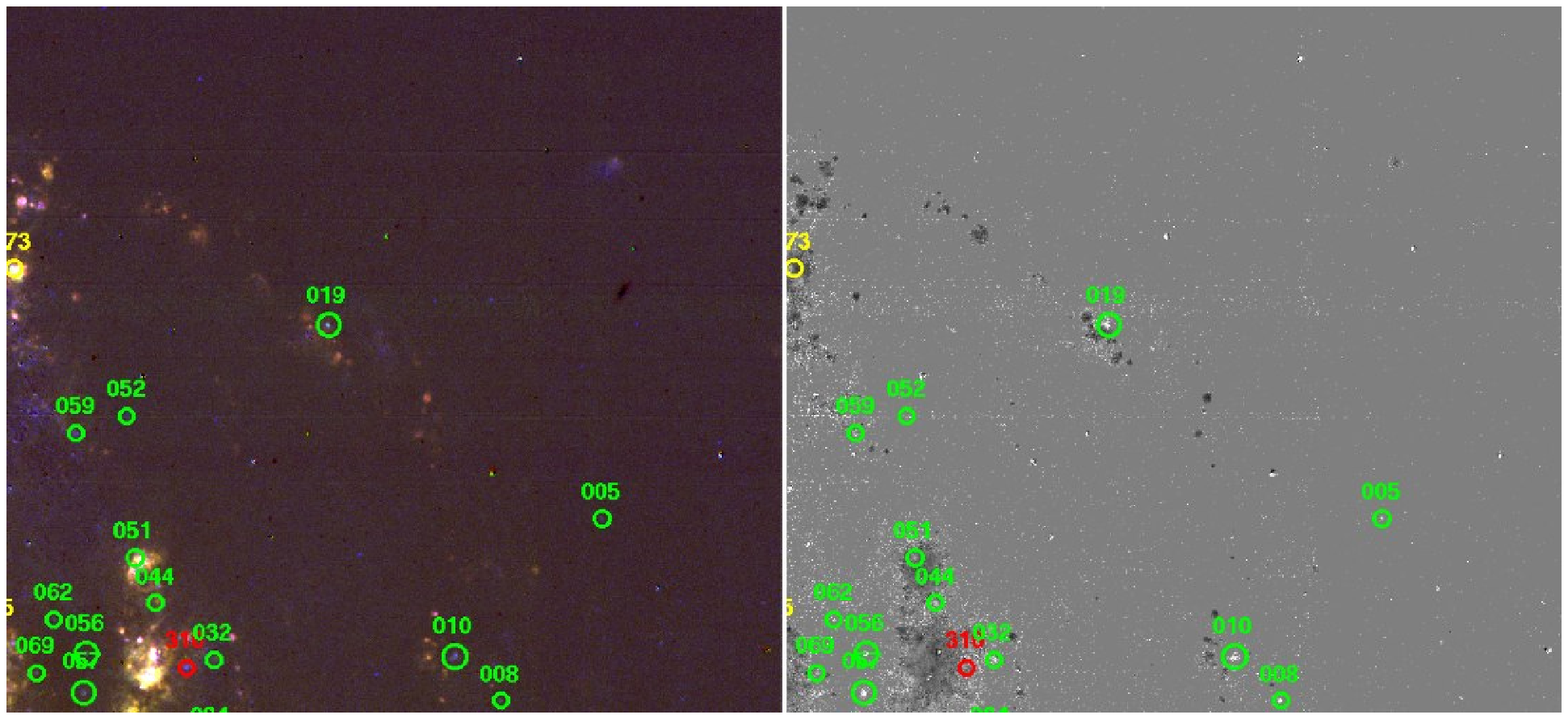}
\caption{Same as Fig.\ \ref{fig_F1} but for Field 3, as indicated in Fig.\ \ref{fig_overview}. \label{fig_F3}}
\end{figure}

\begin{figure}
\plotone{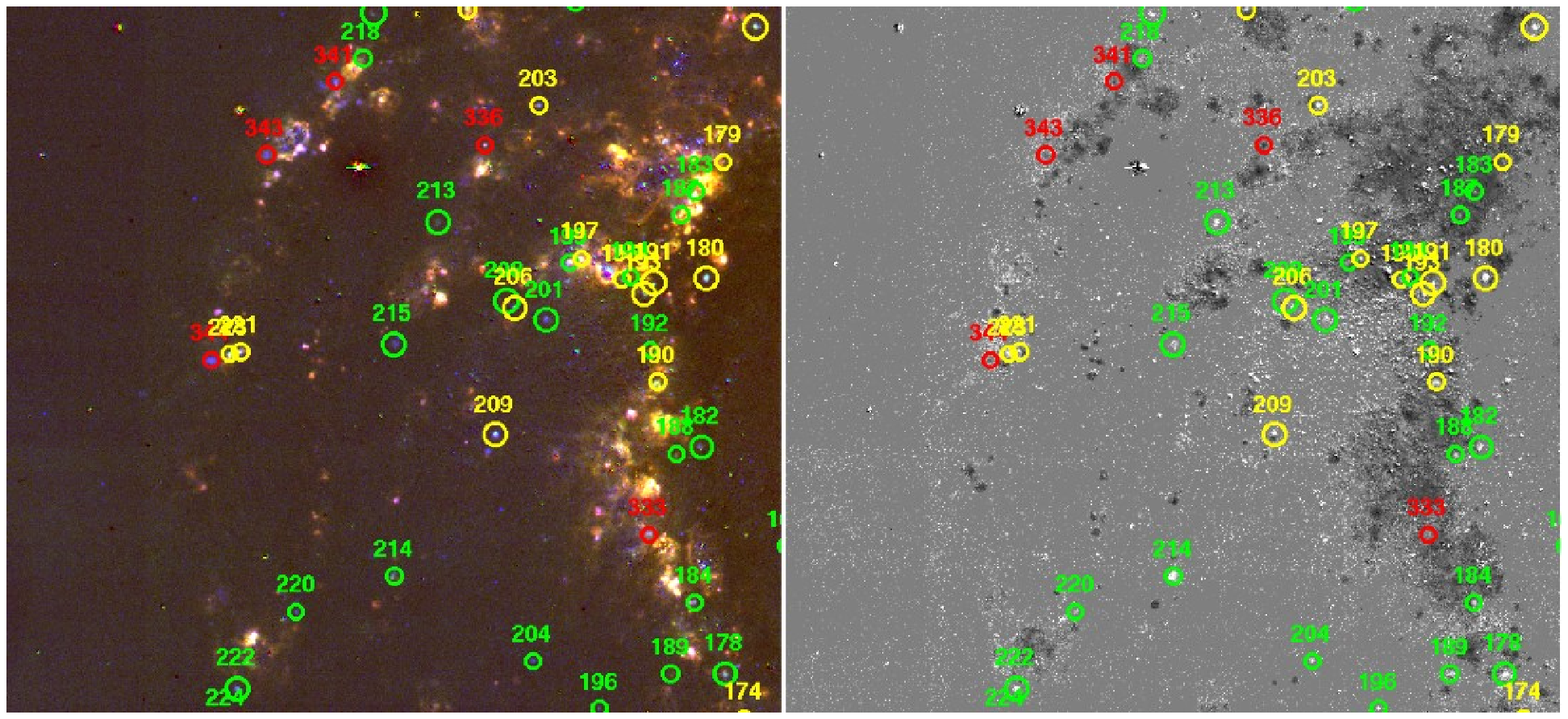}
\caption{Same as Fig.\ \ref{fig_F1} but for Field 4, as indicated in Fig.\ \ref{fig_overview}. \label{fig_F4}}
\end{figure}

\begin{figure}
\plotone{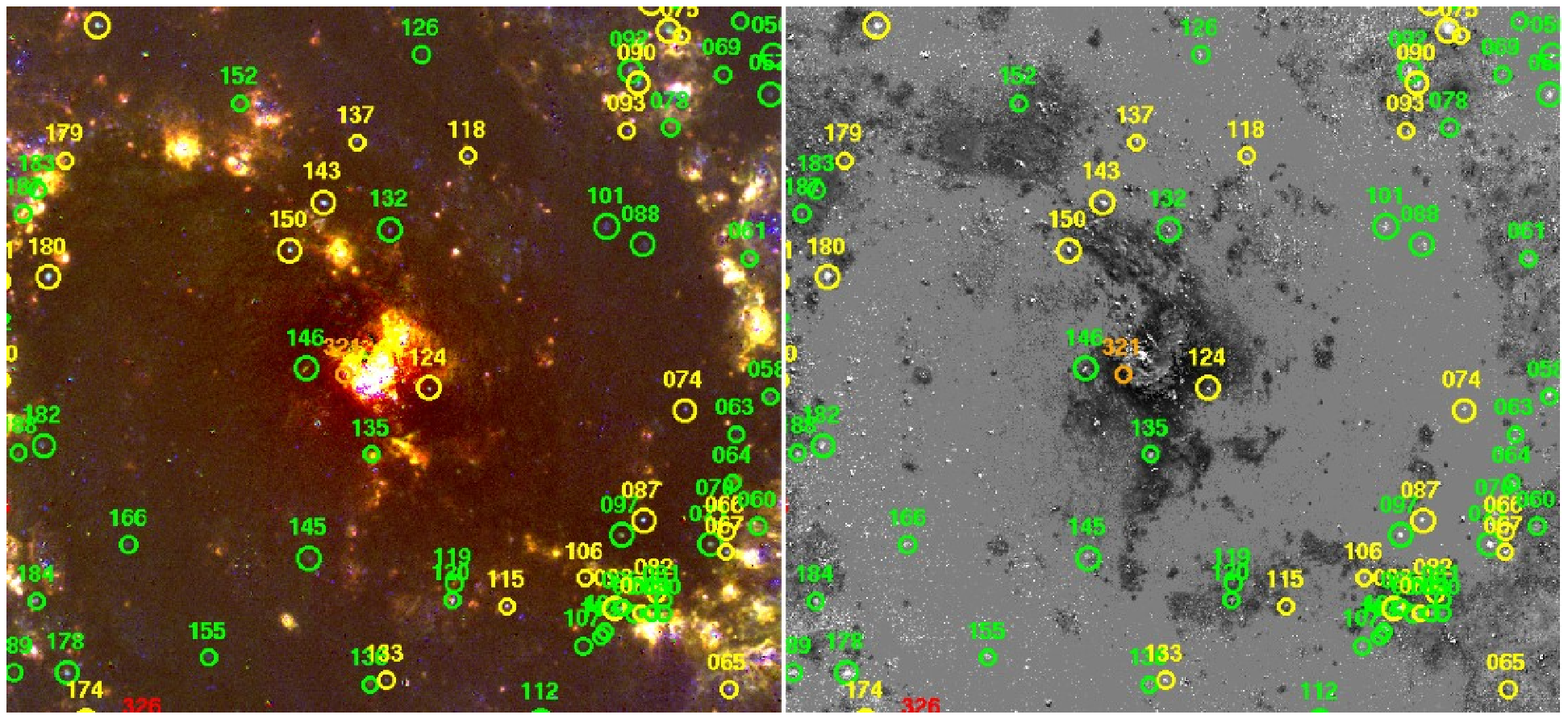}
\caption{Same as Fig.\ \ref{fig_F1} but for Field 5, as indicated in Fig.\ \ref{fig_overview}. \label{fig_F5}}
\end{figure}

\begin{figure}
\plotone{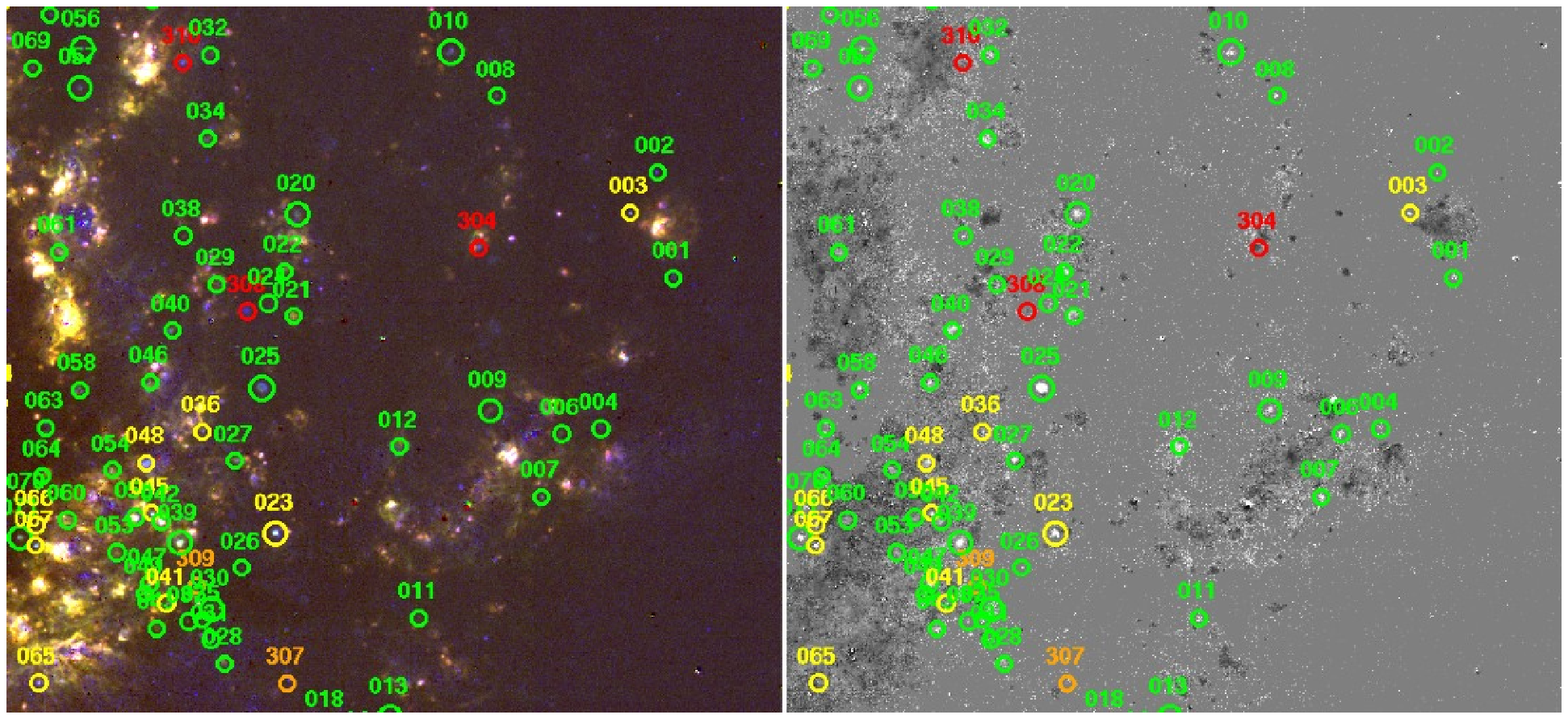}
\caption{Same as Fig.\ \ref{fig_F1} but for Field 6, as indicated in Fig.\ \ref{fig_overview}. \label{fig_F6}}
\end{figure}

\begin{figure}
\plotone{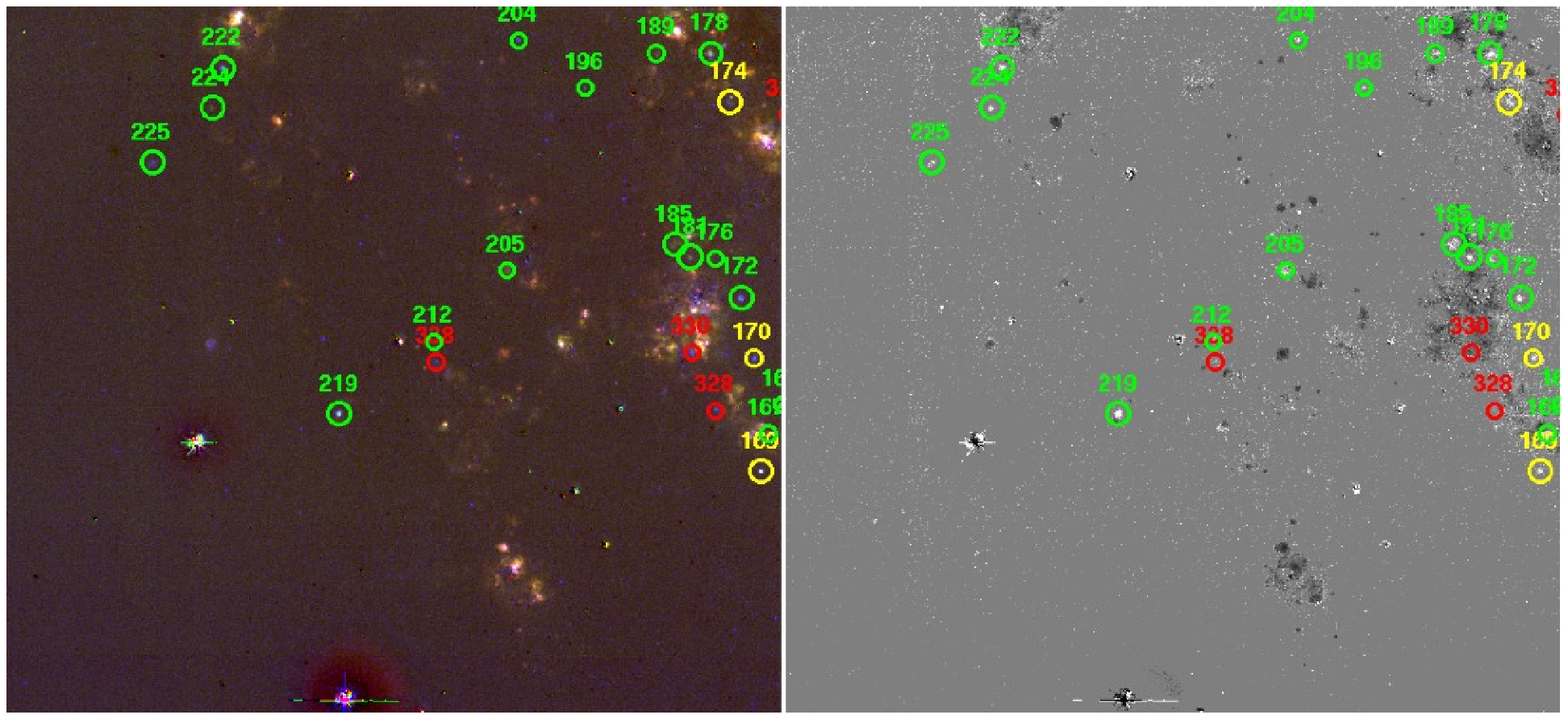}
\caption{Same as Fig.\ \ref{fig_F1} but for Field 7, as indicated in Fig.\ \ref{fig_overview}. \label{fig_F7}}
\end{figure}

\begin{figure}
\plotone{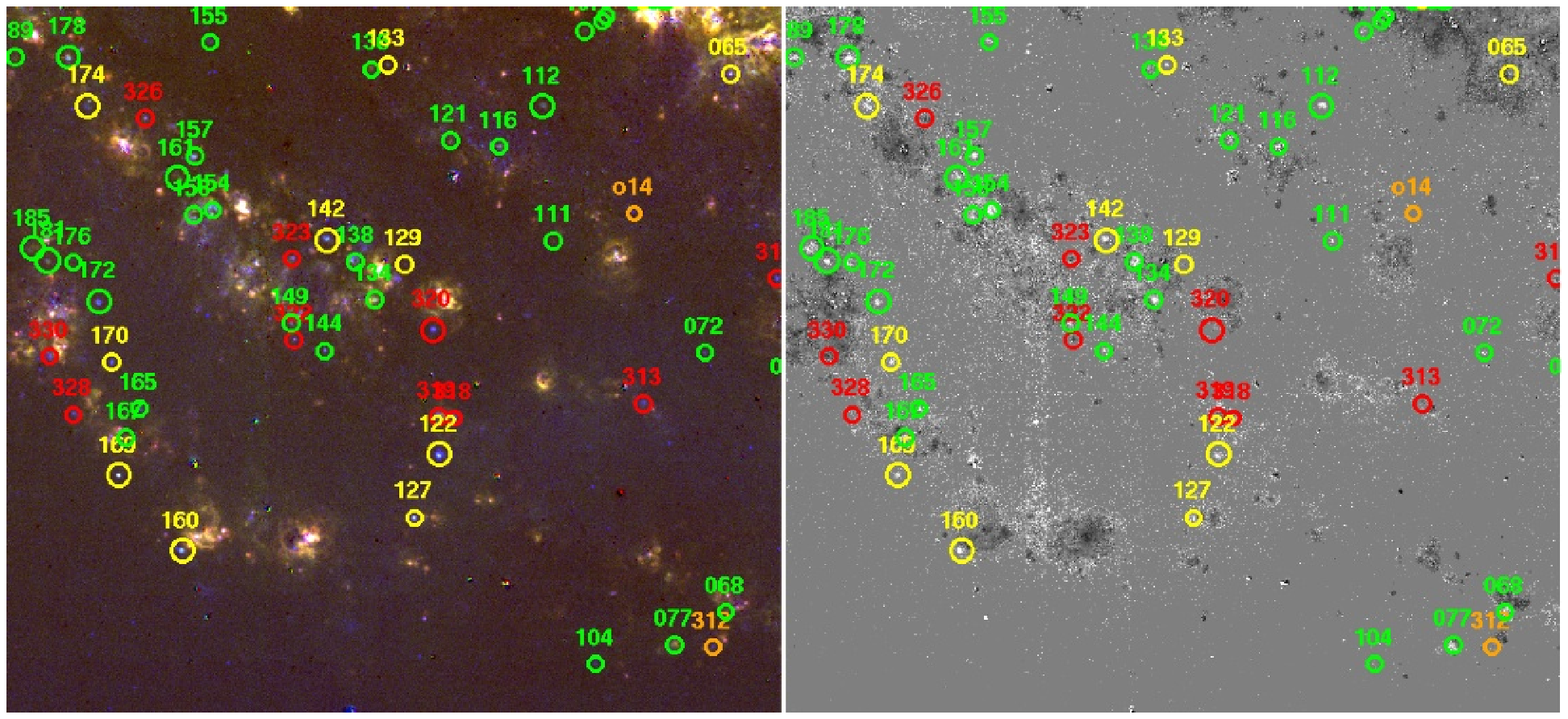}
\caption{Same as Fig.\ \ref{fig_F1} but for Field 8, as indicated in Fig.\ \ref{fig_overview}. \label{fig_F8}}
\end{figure}

\end{document}